\documentclass{pasj02}

\usepackage{amsmath}
\usepackage[switch,mathlines]{lineno} 

\usepackage{siunitx}

\Received{$\langle$reception date$\rangle$}
\Accepted{$\langle$acception date$\rangle$}
\Published{$\langle$publication date$\rangle$}

\begin{document}

\title{Formation of Parallel Stellar Streams through Encounters with Dark Matter Subhalos and Intermediate-Mass Black Holes}

\author{Yuka \textsc{Kaneda}\altaffilmark{1}\thanks{}}
\altaffiltext{1}{Department of Astronomy, The University of Tokyo, Bunkyo-ku, Tokyo 113-0033, Japan}
\email{kaneda@astron.s.u-tokyo.ac.jp}

\author{Masao \textsc{Mori}\altaffilmark{2}\thanks{}}
\altaffiltext{2}{Center for Computational Sciences, University of Tsukuba, Tsukuba, Ibaraki 305-8577, Japan}
\email{mmori@ccs.tsukuba.ac.jp}

\author{Yohei \textsc{Miki}\altaffilmark{3}}
\altaffiltext{3}{Information Technology Center, The University of Tokyo, Kashiwa, Chiba 277-0882, Japan}

\author{Takanobu \textsc{Kirihara}\altaffilmark{4}}
\altaffiltext{4}{Kitami Institute of Technology, 165, Koen-cho, Kitami, Hokkaido 090-8507, Japan}

\author{Andreas \textsc{Burkert}\altaffilmark{5,6}}
\altaffiltext{5}{University Observatory Munich, Scheinerstrasse 1, D-81679 Munich, Germany}
\altaffiltext{6}{Max-Planck-Institut f\"{u}r extraterrestrische Physik, Garching 85748, Germany}


\maketitle

\begin{abstract}
Dark matter subhalos and intermediate-mass black holes wandering in the Milky Way and the Andromeda galaxy are difficult to directly detect through electromagnetic observations, yet knowing their abundance is essential for understanding galaxy formation and evolution. 
We propose parallel stellar streams as dynamical imprints left on stellar streams by dark perturbers, including starless dark matter subhalos and wandering intermediate-mass black holes. 
We report that a single stream can split into two parallel structures after an encounter with a dark perturber. 
This scenario is supported by analytical modelling and $N$-body simulations. 
We also discuss how we can distinguish parallel stellar streams from other formation processes based on observables.
We extend the theoretical picture of stream--subhalo interactions by showing that encounters with dark perturbers can generate density depletions perpendicular to the stream elongation, leading to parallel stellar stream morphologies beyond conventional gap-like signatures.
\end{abstract}


\section{Introduction}
At the luminous end of the galaxy population, the widely accepted view regarding the discrepancy between the galaxy luminosity function and the dark matter halo mass function is that luminous galaxies are less abundant than theoretical predictions, owing to the suppression of star formation by feedback from supermassive black holes.
However, the mechanisms driving supermassive black hole growth remain unclear, partly due to the limited evidence for intermediate-mass black holes (IMBHs, $100-10^5~M_\odot$), which are hypothesized to bridge the mass gap between well-established stellar-mass black holes and supermassive black holes \citep{Abbott2020, Abac2025}.
Gravitational-wave observations have identified IMBHs with masses around $100~M_\odot$ resulting from stellar-mass black hole mergers.
Dwarf satellites have long been considered promising hosts of IMBHs. 
Indeed, recent discoveries of $10^5~M_{\odot}$ black holes in nearby dwarf galaxies have been identified via broad-line active galactic nucleus emission \citep{ReinesVolonteri2015, MezcuaDominguezSanchez2024}. 
The growing numbers at the less-massive and massive end of the IMBH range points to a broader, yet unseen, population in between.
Detecting such IMBHs is particularly challenging in the gas-poor dwarf satellites of the Milky Way (MW) and the Andromeda galaxy (M31), where many are likely quiescent due to the lack of accreting material. Some may have originated at the centers of dwarf galaxies that were later accreted by their hosts, with their stars dispersed into the stellar halo and the black holes left behind as wandering remnants \citep{Kawaguchi+2014, Miki+2014}.

The concordant Cold Dark Matter (CDM) model has been found to overpredict the number of satellite halos around luminous galaxies, such as the MW, compared to observations \citep{Moore1999, Ishiyama2009}.
This tension is alleviated through the latest observational instruments, which increase the number of faint satellite galaxies. 
Meanwhile, in hydrodynamic simulations with a full prescription of baryonic feedback, galaxy formation in low-mass dark matter subhalos is suppressed \citep{D'Onghia2010, Brooks2013, Brooks&Zolotov2014, Sawala2015, Wetzel2016, Applebaum2021ApJ}.
Moreover, baryonic suppression of star formation results in an abundant population of starless dark matter subhalos that remain invisible because they lack the stellar light that makes galaxies observable \citep{Jung2024}.
Interestingly, the population of these starless dark matter halos is strongly related to the particle nature of dark matter, including the particle mass \citep{Lovell2014, Du2018, banik_novel_2021, banik_evidence_2021}.
Therefore, constraining the subhalo mass function in the low-mass regime provides a crucial route to revealing the hidden nature of dark matter.
The findings from these previous studies have shifted the focus of the field from the missing satellite problem to the starless dark matter subhalo detection problem.

Pioneering studies pointed out that stellar streams are sensitive to the lumpy structure of the host dark matter halo \citep{Johnston2002, Ibata2002}.
Subsequent studies pointed out that starless dark matter subhalos can create gaps in stellar streams when they interact with the stream (e.g. \cite{Carlberg2011, yoon_clumpy_2011, Carlberg2012_Pal5, CarlbergGrillmair2013, ErkalBelokurov2015a, ErkalBelokurov2015b, Carlberg2016, Erkal2016, Kirihara2017_NWS, Komiyama2018}). 
Several methods have been proposed to search for starless dark matter subhalos. Strong gravitational lensing provides a powerful probe of dark matter substructure through localized perturbations to lensed images, while indirect searches attempt to identify subhalos via possible dark matter annihilation signals. However, the sensitivity of strong lensing is generally limited to relatively massive subhalos, typically $\gtrsim10^{7-9}\,M_\odot$ depending on the lens configuration \citep{vegetti_bayesian_2009}, and indirect searches have not yet established a robust connection between gamma-ray emission and the expected locations of low-mass subhalos \citep{buckley_dark_2010, vivier_veritas_2011}. In this context, stellar streams provide a complementary and particularly promising probe of low-mass dark matter subhalos through the dynamical imprints induced on the stream morphology and kinematics. Because these signatures arise purely from gravitational interactions, stellar streams offer a unique opportunity to probe the low-mass end of the dark matter subhalo population, which is closely connected to the fundamental nature of dark matter.

Stellar streams have been widely used as sensitive probes of dark matter subhalos, primarily through the identification of density deficits, or gaps, projected along the stream elongation direction (e.g. \cite{Carlberg2011}). Previous studies investigated such density depletions by tracing particle motions mainly along the stream direction using analytical models and simulations (e.g. \cite{ErkalBelokurov2015a}). In contrast, density variations perpendicular to the stream direction have received relatively little attention. To address this, we investigate particle motions perpendicular to the stream elongation and explore the resulting density variations across the stream.

\begin{figure}
	\begin{centering}
	    \includegraphics[width=8cm]{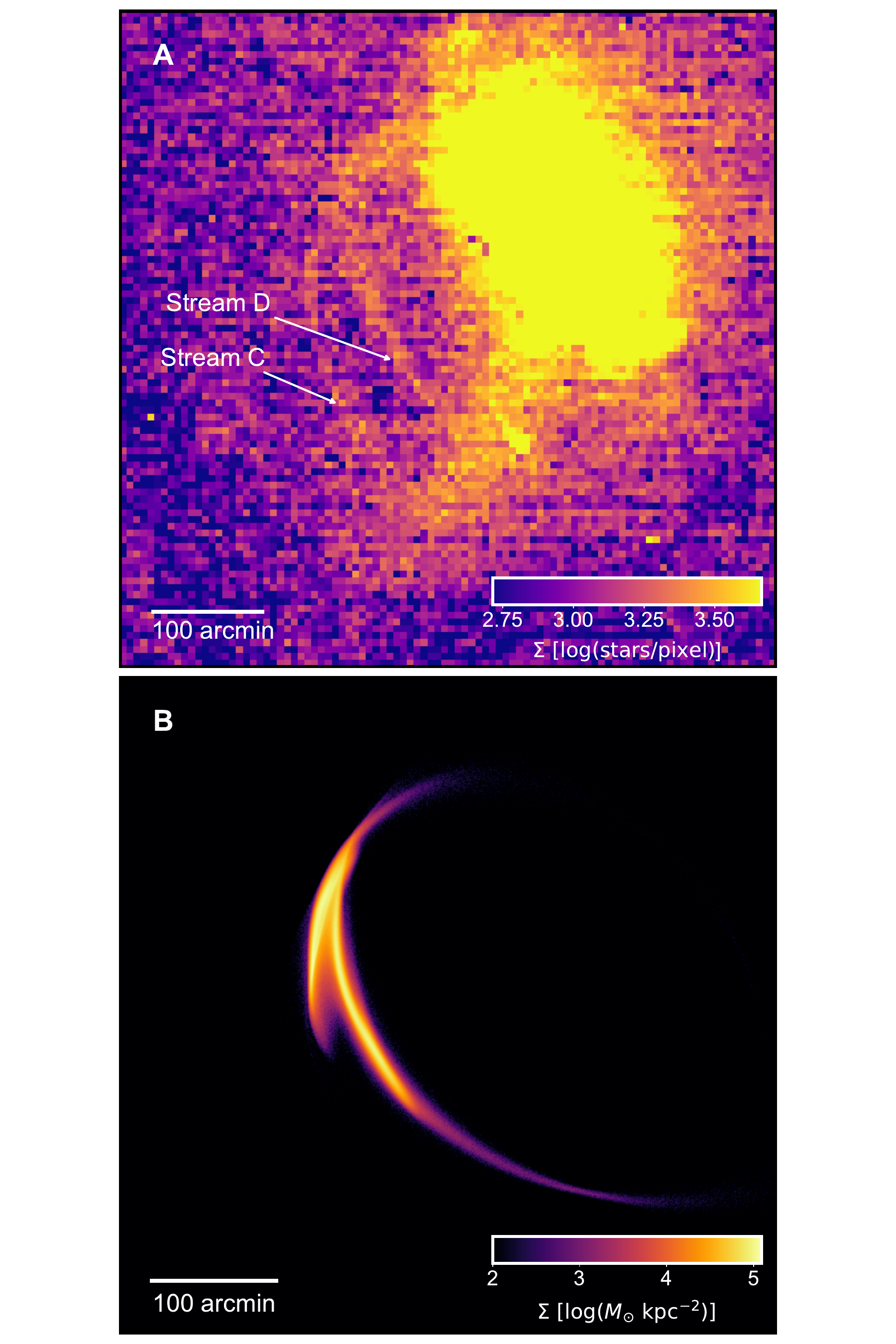}
	\end{centering}
	\caption{   
        Comparison of the star count map in the Andromeda halo (\textbf{A}) and our simulation (\textbf{B}). Panel (\textbf{A}) displays red giant branch star counts in the metallicity range $-1.7 < \text{[Fe/H]} \leq -1.1$ \citep{McConnachie2018}. Stream C and Stream D are marked with white arrows and are observed to be nearly parallel and of similar width \citep{Preston2021}. The M31 disk lies in the upper-right.
        Panel (B) shows the projected density of stellar particles in our simulation, illustrating that a stream--perturber encounter can produce a two-component morphology qualitatively similar to Streams C and D for a suitable viewing angle.
        The scale is converted using the M31 distance from Earth as 783 kpc \citep{McConnachie2018}, which gives ${100}~\mathrm{arcmin} = {23}~\mathrm{kpc}$.
        {Alt text: Panel A: Star count map of the Andromeda halo showing Stream C and Stream D as two nearly parallel stellar streams marked with white arrows. Panel B: Simulated projected stellar particle density showing two parallel stream components produced by a stream-perturber encounter.}
        }
	  \label{fig:comp_obs}
\end{figure}

Studies on the stellar halo of the M31 have revealed substructures \citep{Chapman+2008}, specifically identifying Stream C and Stream D, as illustrated in figure~\ref{fig:comp_obs}A. 
These faint streams, discovered through deep imaging, overlap on the star-count map, orbit at similar distances from the center of the M31, lie at similar heliocentric distances, exhibit comparable widths, and share similar metallicities \citep{conn_major_2016, Preston2021, ogami_structure_2024}.
These features indicate that their progenitor galaxies shared remarkably similar orbits at roughly the same time and dynamical masses.
Moreover, they have experienced nearly identical star formation histories.
These features are difficult to reconcile with the tidal disruption of two distinct dwarf galaxies.
To date, only one theoretical model has been proposed \citep{Fardal2008}, but it adopts a fundamentally different scenario from the one we propose below and struggles to reproduce the complex observational properties recently found in Stream C and Stream D.
Here, we hypothesize that these streams result from the split of a single stellar stream. 
We therefore propose that the structure formed through a collision by a dark object in the parallel--like direction along the stream.
Intriguingly, parallel streams that share the physical properties (e.g. Jhelum and Indus: \cite{shipp2018, Bonaca2019_multiple}) and a stream with multiple internal parallel components (e.g. Jhelum: \cite{Bonaca2019_multiple}) have been observed in the MW,
where numerous stellar streams, many of which originate from tidally disrupted globular clusters, have been identified by the Gaia satellite \citep{Ibata2021,Malhan2022_atlas}. 
In other words, the presence of parallel stellar streams without a luminous collision counterpart may serve as a promising signature of a dark perturber, potentially a starless dark matter subhalo or a wandering IMBH.

Motivated by these situations, this work explores an overlooked regime of stream--perturber interactions by focusing on density variations perpendicular to the stream direction, which can generate parallel stellar stream morphologies. We develop an analytic model that describes density perturbations projected perpendicular to the stream’s elongation. 
Here, please note that the goal of this study is not to reproduce any specific observed stream, but to clarify the physical mechanism for forming parallel stellar streams.

This paper is organized as follows.
In Section 2, we present an analytical description of how a stellar stream can be split by an encounter with a dark matter subhalo or an IMBH.
In Section 3, we test our hypothesis using $N$-body simulations of a stellar stream formed through the tidal disruption of a dwarf galaxy.
In Section 4, we discuss how parallel stellar streams can be distinguished observationally from other formation scenarios using current and upcoming facilities, and present an order-of-magnitude estimate of the expected encounter rate in MW-sized galaxies.
Finally, we summarize our conclusions in Section 5.

\section{Analytical Model for Stream Splitting}

To clarify the underlying physical processes in the formation of parallel stellar streams, we established an analytic model.
The model tracks the trajectories of stream particles around the host potential after receiving an initial velocity kick from the perturber. 
We solve the equations of motion of stream particles up to the first-order terms of the perturbation.
The impulse approximation is adopted to model the perturbation.
For simplicity, the self-gravity of the stream is ignored.
Our analytical model assumes that the encounter takes place in the orbital plane of the stream and that the perturbed stream particles remain on this plane after the encounter.

\begin{figure}
    \begin{center}
        \includegraphics[width=8cm]{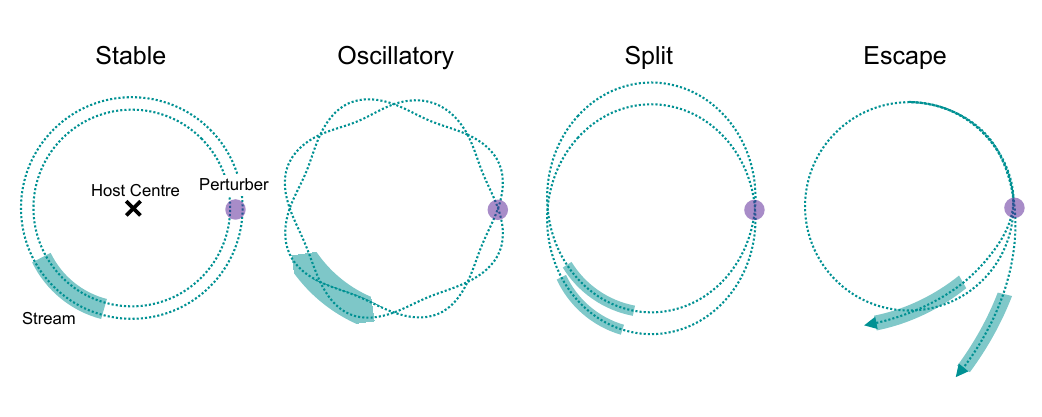}
    \end{center}
    \caption{
     Classification of stream--perturber interactions.
     Stream--perturber interactions can be classified into four distinct cases: \textit{Stable}, \textit{Oscillatory}, \textit{Split}, and \textit{Escape}. In the \textit{Stable} case, the stream remains largely unaffected by the perturber, and no splitting occurs. 
     In the \textit{Oscillatory} case, the perturber causes the stream to broaden; however, the self-gravitational binding of the stream prevents significant splitting. 
     In the \textit{Split} case, the stream is divided into multiple streams due to interactions with the perturber, while orbiting the host galaxy. 
     Finally, in the \textit{Escape} case, the stream escapes the gravitational field of the host galaxy following an intense interaction with the perturber, resulting in substantial changes to its orbit.
     {Alt text: Schematic showing four stream-perturber interaction regimes from left to right: Stable (stream unaffected), Oscillatory (stream broadened but bound), Split (stream divided into parallel components), and Escape (stream unbound from host galaxy).}
    }
    \label{fig:classification}
\end{figure}

Effects with minimal impact on particle trajectories are approximated or ignored to construct a simplified model. 
In the scenario under consideration, the gravitational potentials of the host halo, stream, and perturber determine the stream's behaviour, 
which we classify into four cases based on the dominant potential. Schematic representations are shown in figure~\ref{fig:classification}. 
In the first three cases, the host halo's potential dominates. When the perturber's potential is negligible, the stream's orbit remains unaffected, 
classified as the \textit{Stable} case. If the perturber's influence increases, the stream is perturbed but remains self-bound, 
defining the \textit{Oscillatory} regime. When the perturber's potential exceeds the stream's self-gravity, the stream's orbit deviates significantly, 
classified as the \textit{Split} case. Finally, if the velocity imparted by the perturber surpasses the host halo's escape velocity, 
the stream is destroyed, which we term the \textit{Escape} case. We focus on the \textit{Split} case, 
where observable widths in the stream result from significant perturbation.

\begin{figure}
    \begin{center}
        \includegraphics[width=8cm]{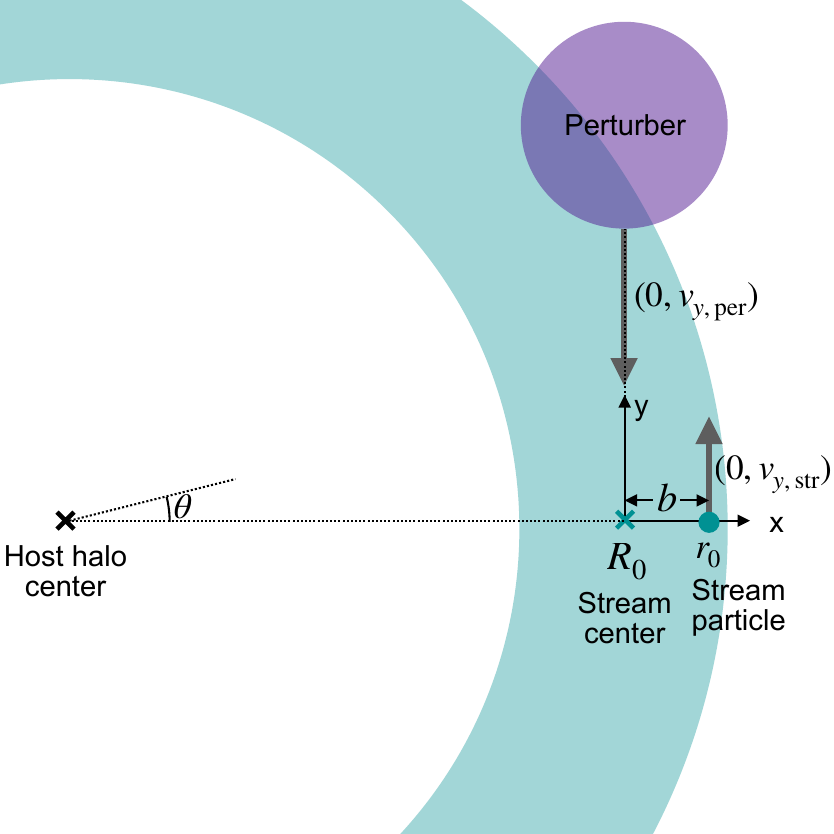}
    \end{center}
    \caption{
    Setup of our analytic model.
    The host halo center lies to the left in the figure.
    The stream has a circular orbit at radius $R_0$ from the host center.
    The axes are stream oriented along the $y$-direction, with $x$ in the radial direction in the host potential.
    We follow the track of a stream particle at (b, 0), which corresponds to the distance $r_0 = R_0 + b$ from the host center, moving in the positive $y$-direction with velocity $(0, \, v_{y, {\rm{str}}})$.
    The perturber is moving with velocity $(0,\, v_{y, {\rm{per}}})$ in the negative $y$ direction.
    The interaction between the stream and the perturber occurs at $(0,\, 0)$. 
    Particles start at $\theta = 0$ and are scattered by the perturber, spreading out into distinct regions.
    {Alt text: Diagram of analytic model geometry showing a stream particle orbiting at radius R-zero from the host halo center, with the perturber approaching and the encounter occurring at the coordinate origin.}
    }
    \label{fig:ana_setup}
\end{figure}

Therefore, our model ignores the self-gravity of the stream and accounts for the host and the perturber potential.
We approximate the stream as an aggregation of massless particles moving under the gravitational potentials of the host halo and the perturber.
The setup for our analytic model is shown in figure~\ref{fig:ana_setup}.
The stream moves in a spherical potential $\phi (r)$, on a circular orbit with radius $R_0$, with velocity $v_{y, {\rm{str}}} = \sqrt{R_0 \partial_r\phi(R_0)}$.
The axes are stream-oriented along the $y$-direction, with $x$ in the radial direction in the host potential.
We consider an interaction of a perturber which is moving in $y$-direction with velocity $(0,\, v_{y, {\rm{per}}})$, and collision occurs at $(0,\, 0)$.
We define the relative velocity between a stream and the perturber along $y$-direction, $v_{\rm{rel}} = v_{y, {\rm{str}}} - v_{y, {\rm{per}}}$.

Next, we compute the velocity change of the stream particle at $(b,\, 0)$ from the passage of a Plummer sphere with mass $M_{\rm{per}}$ and scale radius $r_{\rm{per}}$. 
During the interaction, we assume the perturber moves at a constant velocity.
In the limit where the duration of the encounter is small relative to the crossing time of the system, we can use the impulse approximation to get
\begin{align}
    \Delta v_x 
    & = \int_{-\infty}^{\infty} \frac{G M_{\rm{per}} b}{((v_{\rm{rel}} t)^2 + b^2 + r_{\rm{per}}^2)^{3/2}} \mathrm{d}{t} 
    = \frac{2G M_{\rm{per}} b}{(b^2 + r_{\rm{per}}^2) v_{\rm{rel}}}.
\end{align}

Now that we have the amplitude of the kick $\Delta v_x$, we can compute the resulting orbit of each particle along the stream. 
In what follows, we carry out the analysis at leading order in $\Delta v_x / v_x$ and ignore terms that are $\mathcal{O}\left(\left(\Delta v_x / v_x\right)^2\right)$. 
After the kick, each particle finds itself in a new orbit defined by the kicked initial velocity and unchanged initial position.

We calculate the orbits of stream particles after the kick in a static, spherically symmetric gravitational field of the host halo.
From here on, we take the polar coordinate origin at the host center and set $\theta = 0$ in the direction of the encounter position.
After the initial velocity kick, the angular momentum is conserved:
\begin{align}
    L_z = r_0 v_y = const.
\end{align}
Since we consider particles moving in the orbital plane, we may simply use plane polar coordinates $(r,\, \theta)$ in which the center of the host halo is at $r = 0$ and $\theta$ is the azimuthal angle in the orbital plane.
As the impact occurs at $r = R_0$, the impact parameter is expressed by $b = r_0 - R_0$ in the polar coordinate.
The equation of motion can be simplified by replacing time $t$ by angle $\theta$ and substituting $r$ by $u = 1/r$ as follows:
\begin{align}
   \frac{\mathrm{d}^2{u}}{\mathrm{d}\theta^2} + u=-\frac{1}{L_z^2} \partial_u{\phi}.
\end{align}
This expression is then expanded to the leading order around the original orbit, $u = u_0 + \Delta u$, with $u_0 = 1/r_0$, to get 
\begin{align}
  \frac{\mathrm{d}^2{\Delta u}}{\mathrm{d}\theta^2} + \Delta u = - \frac{1}{L_z^2} \Delta u \left. \partial_u^2{\phi} \right|_{u = u_0}.
\end{align}
When we assume a point-mass potential as $\phi$, the right-hand side of the equation becomes zero.
The solution to the homogeneous differential equation is
\begin{align}
    \Delta u = - u_0 \frac{\Delta v_x}{v_y} \sin \theta, \label{eq:solution_u}
\end{align}
where we have imposed the conditions
$\Delta u (0)  = 0$ and
$\left.\mathrm{d}{u}/{\mathrm{d}\theta}\right|_{\theta = 0} = - u_0 \Delta v_x / v_y$ since the stream particle was initially on a circular orbit and received a velocity kick, $\Delta v_x$, in the radial direction.
Rewriting equation (\ref{eq:solution_u}) in terms of $r = r_0 + \Delta r$, we get
\begin{align}
    r = \frac{r_0}{1 - \frac{\Delta v_x}{v_y} \sin\theta}.   \label{eq:map}
\end{align}

Now we have a map from the positions of particles in the stream at impact to their positions at any later time.
The map between $r_0$ and $r(\theta)$, equation (\ref{eq:map}), allows us to immediately compute the stream density at angle $\theta(t)$. 
If the map is single valued, the conservation of mass is
\begin{align}
    \rho_0 (r_0) r_0 \mathrm{d}{r_0} = \rho(r) r \mathrm{d}{r}.
\end{align}
Note that we only consider the motion of the particle on the orbital plane and the constant angular velocity.
Using this, the density at $r = r(\theta)$ normalized by the initial density $\rho_0(r_0)$ is given by
\begin{align}
    \frac{\rho(r, \theta)}{\rho_0(r_0)} =  \frac{r_0}{r} \left(\frac{\mathrm{d}{r}}{\mathrm{d}r_0}\right)^{-1}.
\end{align}
If the map is not single valued, i.e. once the stream particles pass each other, we must sum over the right-hand side over all $r_0$ which map to $r$.
To plug the map from $r_0$ to $r(\theta)$, equation (\ref{eq:map}), into the density expression we use the following:
\begin{align}
   \frac{\mathrm{d}{r}}{\mathrm{d}r_0}
&= \frac{X - r_0 \frac{\mathrm{d}{X}}{\mathrm{d}r_0}}{X^2} , \label{eq:density} \\
    X &= 1 - \frac{\Delta v_x}{v_y} \sin \theta ,\\
   \frac{\mathrm{d}{X}}{\mathrm{d}r_0}
&= \frac{\mathrm{d}{\Delta v_x}}{\mathrm{d}r_0} \left(- \frac{1}{v_y} \sin \theta\right)
 + \frac{\mathrm{d}{v_y}}{\mathrm{d}r_0} \left(\frac{1}{v_y^2} \sin \theta \right) ,\\
    \frac{\mathrm{d}{\Delta v_x}}{\mathrm{d}r_0}
&= -\frac{2GM_{\mathrm{per}}}{v_{\mathrm{rel}}} \frac{r_{\mathrm{per}}^2-b^2}{(r_{\mathrm{per}}^2+b^2)^2} ,\\
    \frac{\mathrm{d}{v_y}}{\mathrm{d}r_0} &= - \frac{1}{2} \sqrt{\frac{GM{_\mathrm{host}}}{r_0^3}}
    .
\end{align}

\begin{figure}
    \begin{center}
        \includegraphics[width=8cm]{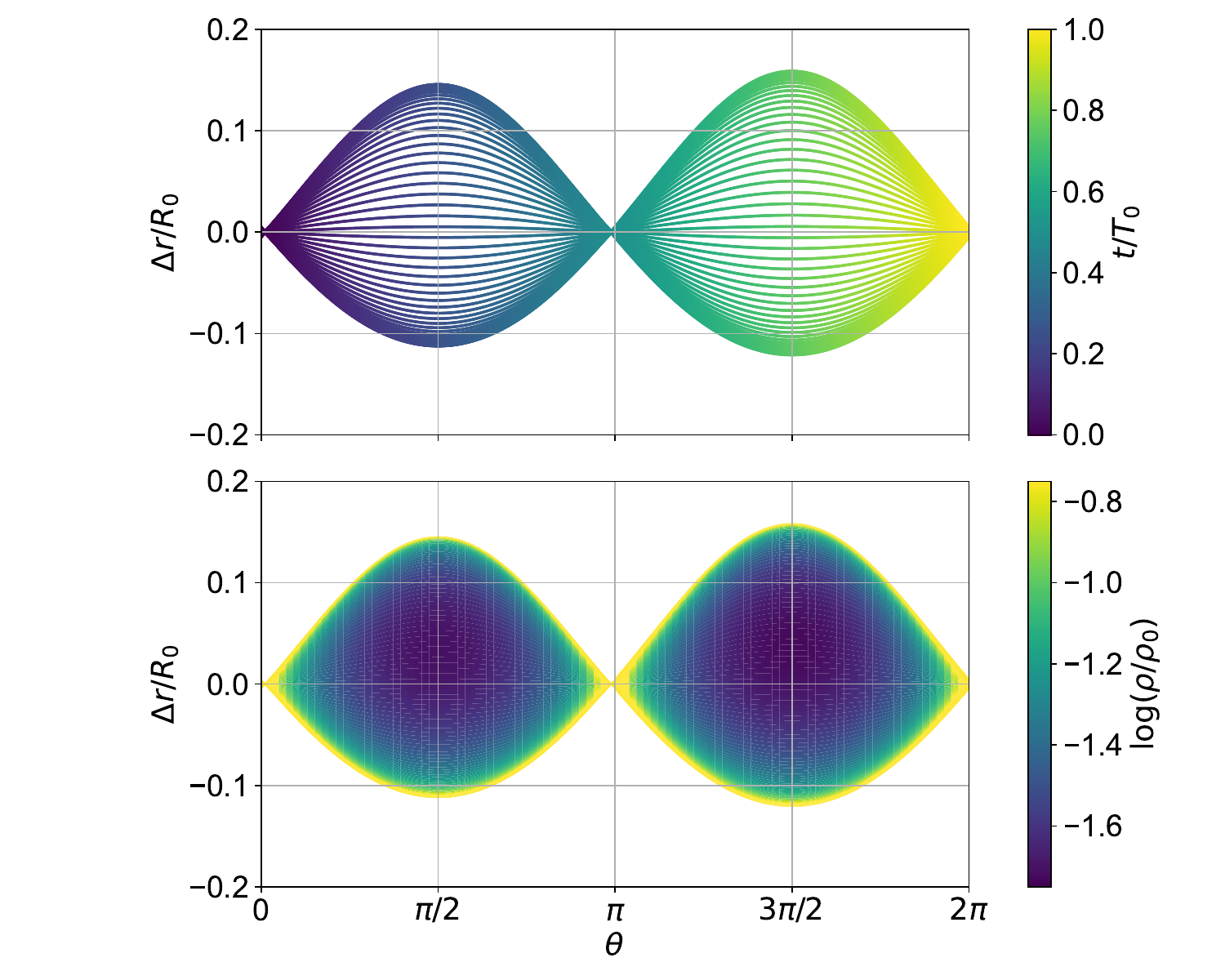}
    \end{center}   
    \caption{
    The track and density distribution of stream particles calculated by the analytical model for the Plummer sphere perturber of $5\times10^8~{M_\odot}$, the core radius {0.32}~{kpc}, and relative velocity ${190}~\mathrm{km~s^{-1}}$.
    Interaction occurs at {60}~{kpc} from the host halo center.
    The vertical axis represents $\Delta r / R_0$, the deviation from the initial stream center normalized by the orbital radius.
    Here, $R_0$ represents the stream center and $\Delta r = r - R_0$ expresses the deviation from the stream center. 
    Particles depart from $\theta = 0$ and are scattered by the perturber, spreading out into distinct regions. 
    The top panel shows the tracks of uniformly sampled stream particles. 
    The color map shows the time dependence normalized by the orbital period $T_0$ at $R_0$.
    The bottom panel shows the logarithmic normalized density of the stream as a color map. 
    {Alt text: Top panel: Stream particle tracks color-coded by orbital period fraction, showing divergence after a Plummer sphere encounter. Bottom panel: Logarithmic normalized density showing two high-density peaks at the stream edges with a depleted central region.}
    }
    \label{fig:ana_density-track_ps}
\end{figure}

Figure~\ref{fig:ana_density-track_ps} shows the results for the Plummer sphere perturber of ${5\times10^8}~{M_\odot}$, the core radius \qty{0.32}{kpc}, and relative velocity ${190}~\mathrm{km\,s^{-1}}$.
Interaction occurs at \qty{60}{kpc} from the host center.
The top panel and the bottom panel in figure~\ref{fig:ana_density-track_ps} show the tracks of uniformly sampled stream particles and the normalized density of the stream after the interaction, respectively. 
This situation corresponds to the scale of the dwarf galaxy stream-dark satellite interaction.
Since the perturber has a mass distribution, the enclosed mass becomes larger when the radius becomes larger.
Therefore, scattering is weaker for particles with small impact parameters and stronger scattering for outer particles.
This results in the convergence of the track of stream particles.
Since the angular momentum is conserved around the host halo, the stream particles cannot exceed certain radius both inward and outward.
These situations cause the edges of the distribution to have the highest density, which is an order of magnitude greater than the lowest-density regions around $\Delta r \sim 0$.
It produces a separation of approximately 15 kpc, which is comparable to the parallel Stream C and Stream D \citep{conn_major_2016, Preston2021, ogami_structure_2024}.
At $\theta = \pi$, the inner and outer positions of the stream particles scattered inward and outward by the initial kick are swapped. 
This is because each kicked stream particle follows an elliptical orbit with one focal point on the host halo center and the other on shifted points.
Here, it is important to note that these plots show the track of stream particles that exceed the limitations of the analytic model, which solves linear equations for perturbation.

\begin{figure}
    \begin{center}
        \includegraphics[width=8cm]{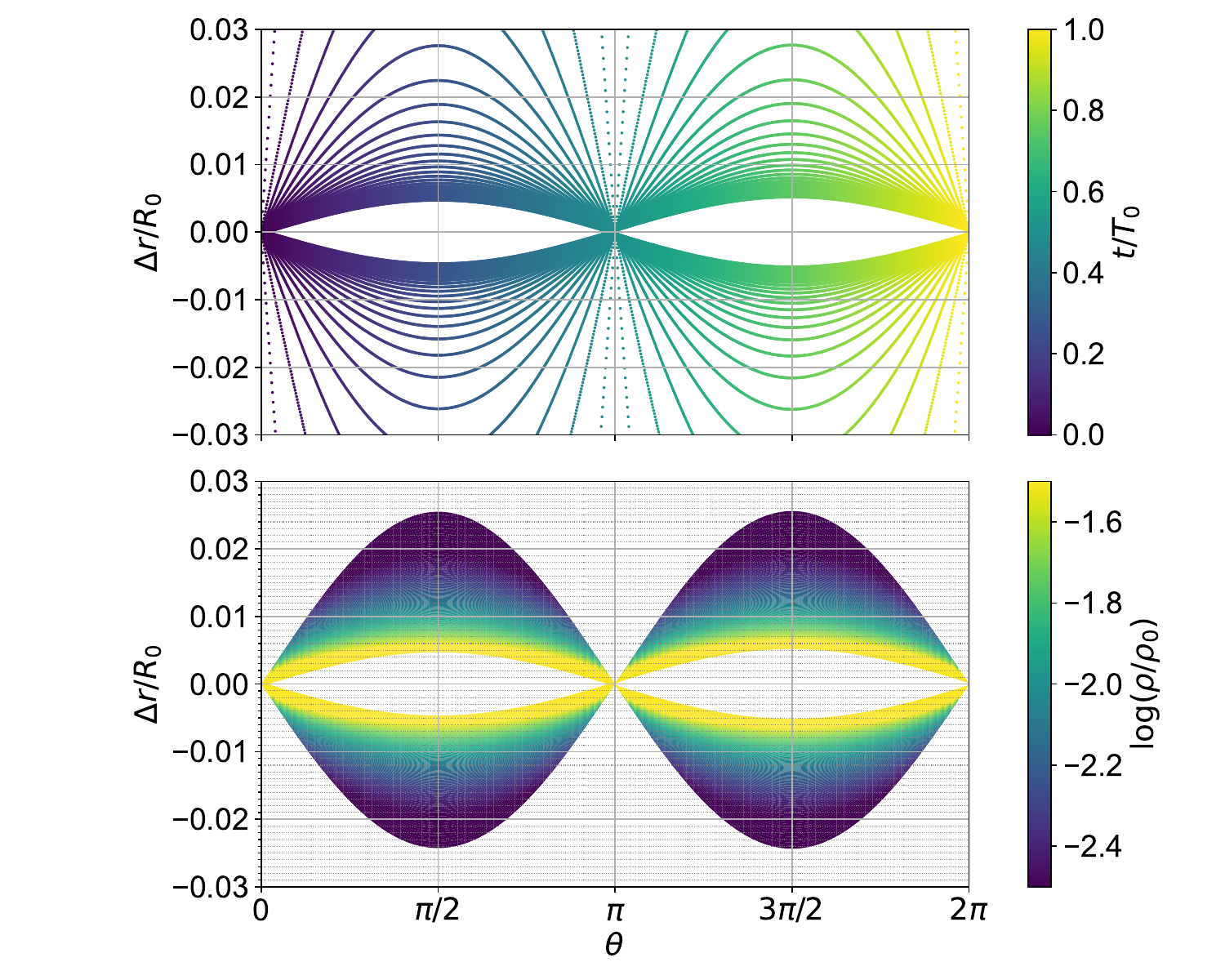}
    \end{center}
    \caption{
    The results of the analytical model for the point mass perturber of ${10^5}~{M_\odot}$, and relative velocity ${70}~\mathrm{km~s^{-1}}$.
    Interaction occurs at {25}~{kpc} from the host center.
    The counterpart stream has a cutoff radius at {6}~{pc} due to the limited size of a globular cluster stream.
    The vertical axis represents $\Delta r / R_0$, the deviation from the initial stream center normalized by the orbital radius.
    Here, $R_0$ represents the stream center and $\Delta r = r - R_0$ expresses the deviation from the stream center. 
    Particles depart from $\theta = 0$ and are scattered by the perturber, spreading out into distinct regions. 
    The top panel shows the tracks of uniformly sampled stream particles. 
    The color map shows the time dependence normalized by the orbital period $T_0$ at $R_0$.
    The bottom panel shows the logarithmic normalized density of the stream as a color map. 
    {Alt text: Top panel: Stream particle tracks showing inner-edge density enhancement after an intermediate-mass black hole encounter. Bottom panel: Logarithmic normalized density showing high-density inner edges and a central gap due to the initial stream width.}
    }
    \label{fig:ana_density-track_pm}
\end{figure}

Figure~\ref{fig:ana_density-track_pm} shows the results for the point mass perturber of ${10^5}~{M_\odot}$ resembling an IMBH, and relative velocity ${70}~\mathrm{km\,s^{-1}}$.
Interaction occurs at \qty{25}{kpc} from the host center.
This situation corresponds to the scale of globular cluster stream-IMBH interaction.
The counterpart stream has a cutoff radius at \qty{6}{pc} due to the limited size of a globular cluster stream.
The top panel of figure~\ref{fig:ana_density-track_pm} shows the track of particles and the normalized density of a stream interacting with a point-mass perturber resembling an IMBH, respectively. 
A point mass causes weaker scattering for particles with larger impact parameters and stronger scattering for inner particles.
Therefore, the stream particles that had a weak scattering stay inside, leading to enhanced densities at the inner edges
A region with no stream particle formed in the center because of the initial stream width, which determines the smallest scattering angle.
Under favorable conditions, it produces a separation of approximately 200 pc, which is comparable to the separation between components in the Jhelum stream \citep{Bonaca2019_multiple}.

\begin{figure}
    \begin{center}
        \includegraphics[width=8cm]{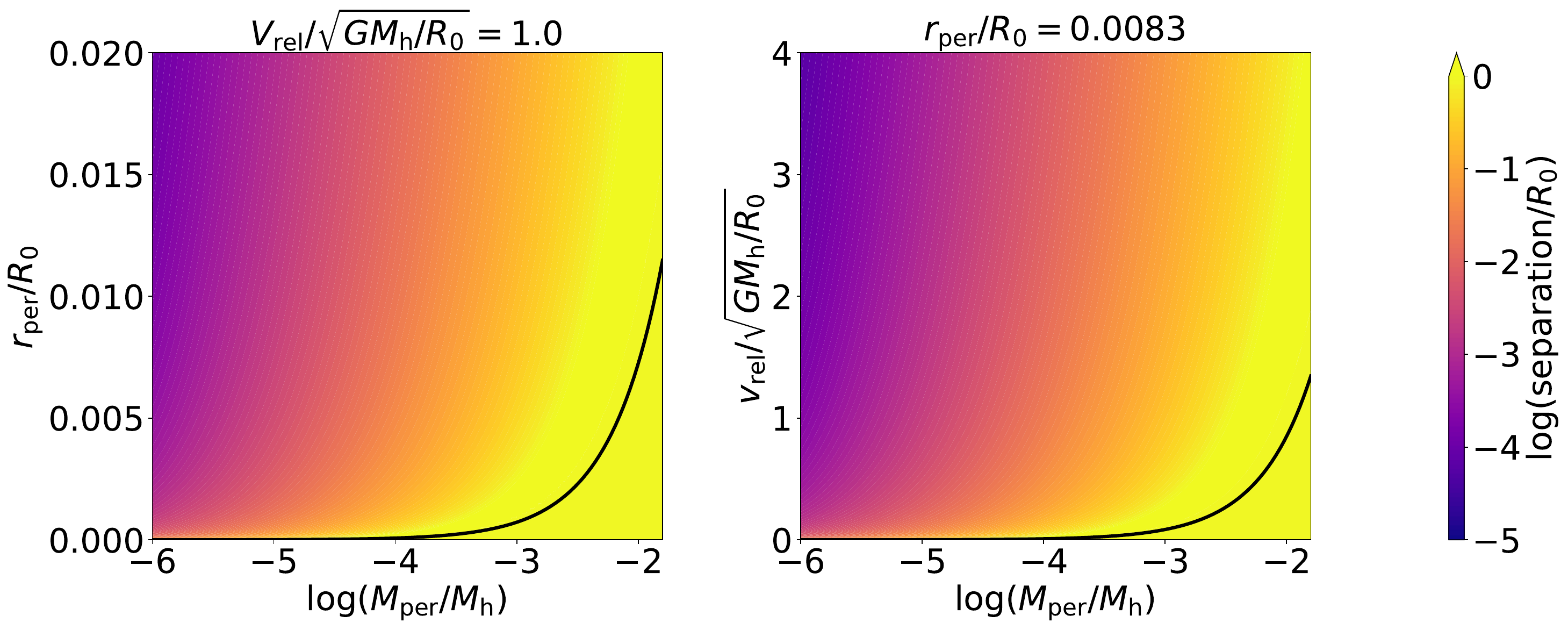}
    \end{center}
    \caption{
    The dependence of maximum stream separation on the mass ratio, size ratio of the perturber to the host, and the relative velocity normalized by the circular velocity of the host halo is shown. 
    The dependence of the maximum separation distance of the parallel streams on the mass ratio and the radius ratio of the perturber and the host halo, and the dependence of the separation distance on the mass ratio and the normalized relative velocity are shown in the left panel and the right panel, respectively.
    Under the black solid line in the bottom panels, the velocity kick exceeds the circular velocity of the host, defining the limit of the analytic model.
    {Alt text: Two color maps of maximum stream separation normalized by orbital radius. Left: dependence on perturber-to-host mass ratio and size ratio. Right: dependence on mass ratio and normalized relative velocity. }
    }
    \label{fig:analyticmodel_density_separation}
\end{figure}

Figure~\ref{fig:analyticmodel_density_separation} illustrates the dependence of maximum stream separation on the mass ratio, size ratio of the perturber to the host, and the relative velocity normalized by the circular velocity of the host.  
The dependence of the maximum separation distance of the parallel streams on the mass ratio and the radius ratio of the perturber and the host halo, the separation distance dependence on the mass ratio and the normalized relative velocity are shown in the left panel and the right panel, respectively.
For instance, when a perturber with a mass of $5\times 10^8~{M_\odot}$ and size of 0.6 kpc collides with the stream that orbits at a distance of $60$ kpc from the center of the host dark matter halo of $10^{12}~{M_\odot}$ at a relative velocity of approximately $\sim 270~\mathrm{km~s^{-1}}$, corresponding to the circular orbital velocity of the host dark matter halo, it causes a separation of 5 kpc.
Greater separations arise from heavier perturbers, more compact sizes, and slower relative velocities, reflecting a deeper gravitational potential, a steeper potential gradient, and a longer interaction timescale, respectively.

\section{$N$-body simulation} \label{sec:simulation}
Inspired by features like Stream C and Stream D, we investigate how dark matter subhalos gravitationally perturb stellar streams. Here, $N$-body simulations reveal that such encounters can split a stellar stream into parallel components, providing insight into the underlying dynamics and their implications for detecting dark perturbers.
However, encounters between IMBHs and stellar streams may require higher-resolution simulations to be modelled accurately. We therefore defer a detailed investigation of this case to future work.

\subsection{Simulation Setup}
The stream is represented by $10^7$ particles with the Plummer softening $\epsilon = \qty{75}{pc}$.
We use the leapfrog method for time integration, the Barnes-Hut tree algorithm with opening angle 0.6 for gravity calculation, and Framework for Developing Particle Simulator (FDPS, \cite{Iwasawa2016}) for parallelisation.
The code solves the evolution of a stream within a host dark matter halo potential and a perturber potential.
The equation of motion of $i$-th stream particle is
\begin{align}
    \frac{\mathrm{d}^2{\boldsymbol{r}_i}}{\mathrm{d}{t}^2}
    = -\sum_{j \neq i} \frac{G m_j\left(\boldsymbol{r}_i-\boldsymbol{r}_j\right)}{\left(\left|\boldsymbol{r}_i-\boldsymbol{r}_j\right|^2+\epsilon^2\right)^{3/2}} - \nabla \Phi_{\mathrm{host}} - \nabla \Phi_{\mathrm{per}},
\end{align}
where
\begin{align}
    \Phi_{\mathrm{host}} = -4\pi\rho_{\rm{s}}r_{\rm{s}}^2 \frac{\ln(1+r/r_{\rm{s}})}{r/r_{\rm{s}}},
\end{align}
is the host potential with Navarro-Frenk-White (NFW)  potential \citep{Navarro1996} and
\begin{align}
    \Phi_{\mathrm{per}} = -\frac{GM_{\rm{per}}}{\sqrt{r^2 + r_{\rm{per}}^2}}
\end{align}
is the perturber potential (Plummer potential)
$\rho_{\rm{s}}$ and $r_{\rm{s}}$ are the scale density and the scale radius of the NFW profile, respectively.
$M_{\rm{per}}$ and $r_{\rm{per}}$ are the core mass and core radius of the Plummer sphere, respectively.
The host halo is modeled to be comparable to the M31,  assuming a total mass of $10^{12}~M_\odot$ and concentration $c_{200} = 10$, and treated as a fixed potential.
For the stream progenitor, the total mass is ${10^7}~{M_\odot}$, which corresponds to the stellar mass as we assume the outer dark matter is already tidally stripped, and the core radius is \qty{0.75}{kpc}.
We adopt a relatively large core radius for the stream progenitor so that the tidal force from the host halo is comparable to the self-gravity of the progenitor, and then the progenitor can form a stream.
Although modelling the perturber as a live dark matter subhalo would be more realistic, such a calculation is computationally expensive to conduct a parameter survey, and so we instead model a perturber as a single particle.
To make a comparison of the stream's response to the perturber's mass, size, and relative velocity, it is desirable for other conditions to remain consistent when we change certain parameters.
Therefore, we assume a perturber with constant velocity for simplicity, while following external potentials such as a stream potential and a host potential is physically correct.
This approximation is reasonable given that the collision timescale is much shorter than the orbital timescale of the perturber.

\subsection{Time Evolution}
\begin{figure}
	\begin{center}
	    \includegraphics[width=8cm]{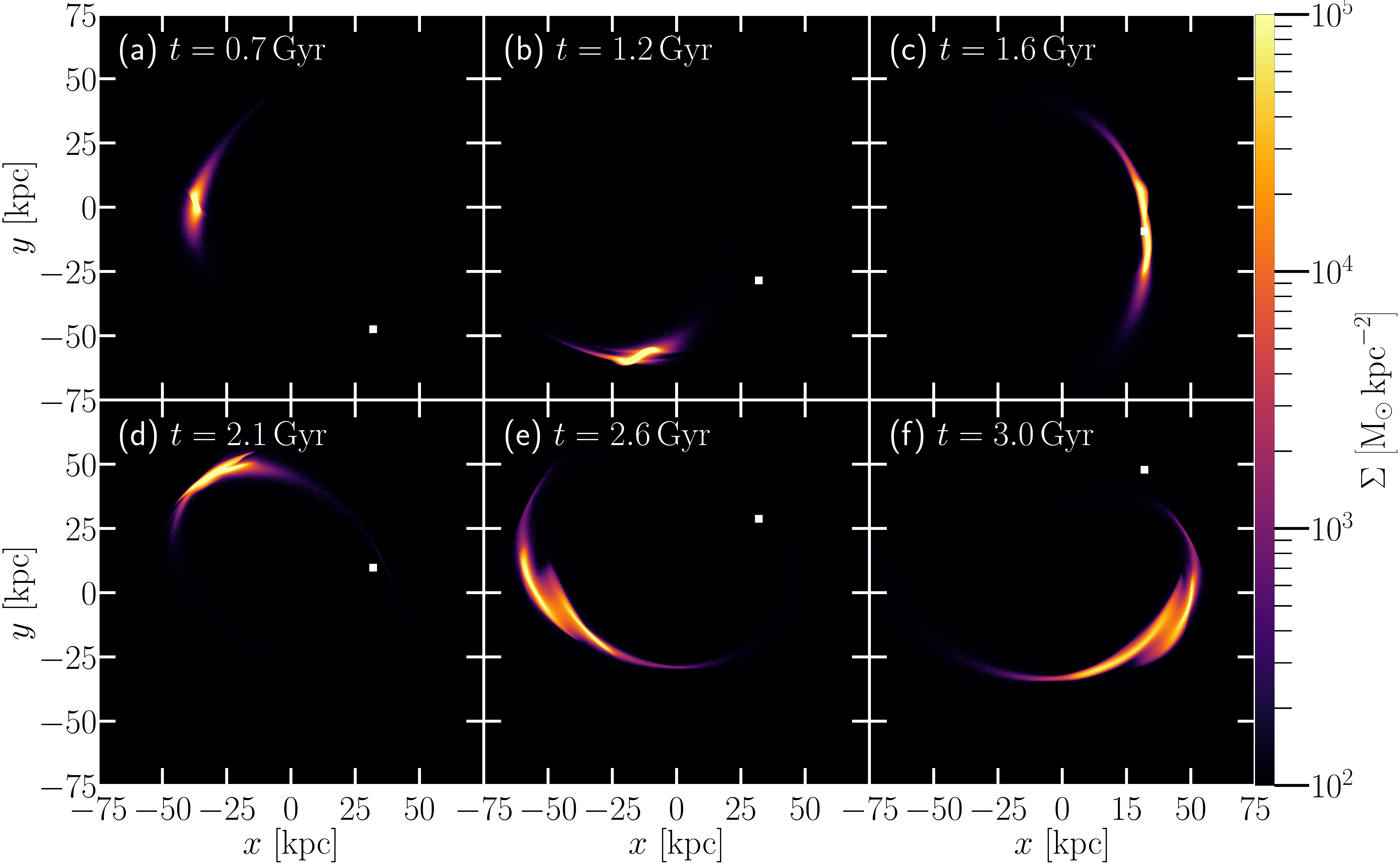}
	\end{center}
	\caption{
    The time evolution of the face-on stream density to the orbital plane.
    The panels labeled a to f show the time evolution of the simulation after the collision between a starless dark matter subhalo and a stellar stream.
    The surface mass density of the stream is shown in the color map, with the mass center of the starless dark matter subhalo indicated by a white dot. In this model, the host dark matter halo has a mass of $10^{12}~{M_\odot}$. The stream progenitor has a mass of ${10^7}~{M_\odot}$, and the starless dark matter subhalo has a mass of ${5\times10^8}~{M_\odot}$, moving at relative velocity ${190}~\mathrm{km\,s^{-1}}$. The time elapsed since the start of the calculation is represented in each panel as follows: (\textbf{a}) 0.7 Gyr, (\textbf{b}) 1.2 Gyr, (\textbf{c}) 1.6 Gyr, (\textbf{d}) 2.1 Gyr, (\textbf{e}) 2.6 Gyr, and (\textbf{f}) 3.0 Gyr.
    {Alt text: Six-panel time sequence from 0.7 to 3.0 gigayears showing face-on stream density evolving from a single arc to two parallel components after an encounter with a dark matter subhalo near 1.6 gigayears.}
    }
	  \label{fig:timeevolution}
\end{figure}

Figure~\ref{fig:timeevolution}a-f illustrates the time evolution of a collision between a starless dark matter subhalo and a stellar stream. 
The stream follows a slightly elongated orbit so as to produce a morphology comparable to that of Stream C and Stream D.
The colliding dark satellite has a total mass of $5\times10^8~M_\odot$, which corresponds to the dark matter mass, as we assume the system is dark matter dominated \citep{Benitez-Llambay2020}, a size of \qty{0.32}{kpc}, and moves at a relative velocity of $190~\mathrm{km\,s^{-1}}$ with respect to the stream.
The surface mass density of the stream is depicted in the color map, and the white dot indicates the mass center of the starless dark matter subhalo.
The host dark matter halo is not visualized.
After \qty{1.6}{Gyr} from the start of the simulation (figure~\ref{fig:timeevolution}c), the perturber collides with the center of the stream and begins to pass through it. By \qty{2}{Gyr} after the start of the simulation (figure~\ref{fig:timeevolution}d), the perturber has fully traversed the stream. Following this interaction, the gravitational scattering induced by the perturber causes particles in the stream to separate into distinct structures on either side. Over time, these structures evolve into well-defined parallel streams (figure~\ref{fig:timeevolution}f). 

A snapshot of the simulation provides an example of a two-component stream morphology that resembles Stream~C and Stream~D in projection.
Panels~A and~B in figure~\ref{fig:comp_obs} compare observations of the southeastern region of the M31 halo with our simulated model. Panel~A illustrates the distribution of red giant branch stars within the metallicity range $-1.7 < \text{[Fe/H]} \leq -1.1$ \citep{McConnachie2018}. Two nearly parallel stellar streams, labeled Stream~C and Stream~D, are highlighted by white arrows. Panel~B presents the projected particle density from our simulation of the colliding system, showing that a stream--perturber encounter can produce a two-component morphology qualitatively similar to Stream~C and Stream~D for a suitable viewing angle.
This simulation suggests that gravitational interactions with a perturber could potentially account for the formation of such parallel stream structures.
However, we emphasize that the goal of this study is not to reproduce any specific observed
stream in detail, but to clarify the physical conditions under which parallel stellar streams can
form and to demonstrate how such structures can be used to constrain the properties of dark
matter subhalos.

\begin{figure}
    \begin{center}
        \includegraphics[width=8cm]{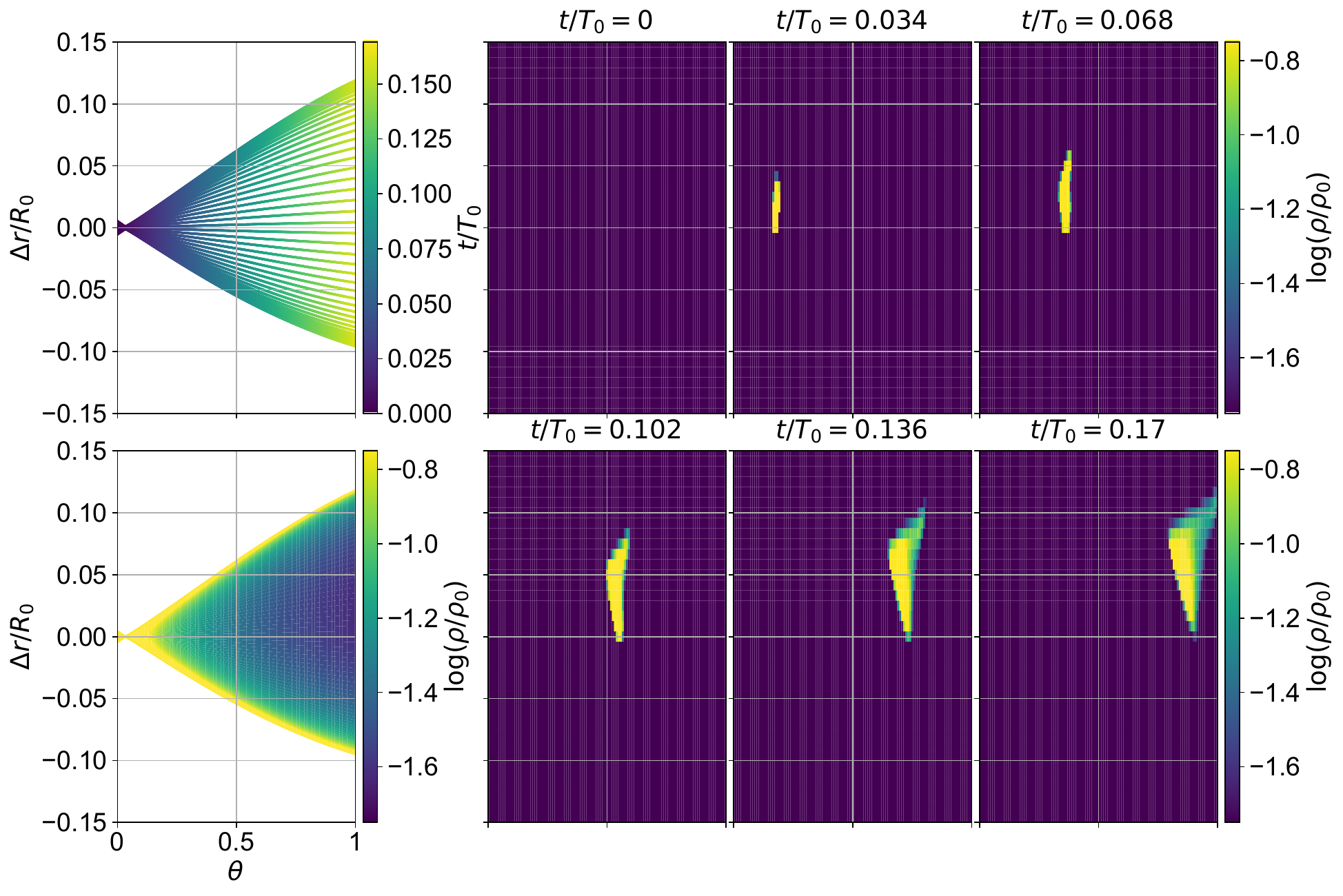}
    \end{center}   
    \caption{Comparison between the analytic model and the simulation.
    Left two panels show the time evolution of the tracks and density distribution of stream particles calculated by the analytic model for the Plummer sphere perturber of $5\times10^8~{M_\odot}$, the core radius {0.32}~{kpc}, and relative velocity ${190}~\mathrm{km~s^{-1}}$.
    Interaction occurs at {60}~{kpc} from the host center.
    The tracks start at $t/T_0 = 0$ with $\theta = 0$ and end at $t/T_0 = 0.17$ with $\theta = 1$, as shown in the color bar in the top left panel.
    The right six panels show the time evolution of the simulation snapshots for the same initial conditions, from top left to bottom right.
    The corresponding time is on the panels. 
    The ranges of the plots and the color bars are matched with the analytic model plots.
    {Alt text: Eight panels comparing analytic model predictions and N-body simulation snapshots of stream particle tracks and density distributions at matched times.}
    }
    \label{fig:ana_simu}
\end{figure}

The analytical model provides an order-of-magnitude description of the expansion of the stream seen in the simulation.
Figure~\ref{fig:ana_simu} compares the analytical model with the corresponding $N$-body simulation. 
Since the analytic model relies on a linear approximation, we restrict the comparison to the region satisfying
$\frac{|\Delta r|}{|R_0|} < \frac{1}{10}$,
and display only this regime. 
The left two panels show the time evolution of the particle trajectories and density distribution predicted by the analytical model for a Plummer-sphere perturber with mass $5\times10^8~M_\odot$, core radius $0.32~\mathrm{kpc}$, and relative velocity $190~\mathrm{km~s^{-1}}$. 
The encounter occurs at a galactocentric distance of $60~\mathrm{kpc}$.
The trajectories start from $t/T_0=0$ at $\theta=0$ and end at $t/T_0=0.17$ at $\theta=1$, as indicated by the color bar in the upper-left panel. 
To compare the simulation results with the analytic model, we follow the following procedure.
First, we follow the particles located in the vicinity of the perturber at the reference time, which we define as $t=0$ and $\theta =0$. 
Second, the density distribution is evaluated in a curvilinear coordinate system, in which one of the coordinate axes is aligned with the orbit of the test particle which represents the stream progenitor. 
The right six panels show the time evolution of the corresponding simulation snapshots for the same initial conditions, from the upper-left to the lower-right panel. 
The time of each snapshot is indicated on each panel. 
The plotting ranges and color bars are matched to those of the analytical-model panels.
In the simulation panels, the center of the stream is slightly offset because we choose the orbit of the test particle as one of the coordinate axes, which is not strictly exact. 
Nevertheless, the stream width predicted by the analytical model agrees with that measured in the simulation to the order of magnitude.

\subsection{Phase Space Behavior}
In this section, we will explain why the stream splitting is most significant in the orbital plane.

\begin{figure}
    \begin{center}
        \includegraphics[width=8cm]{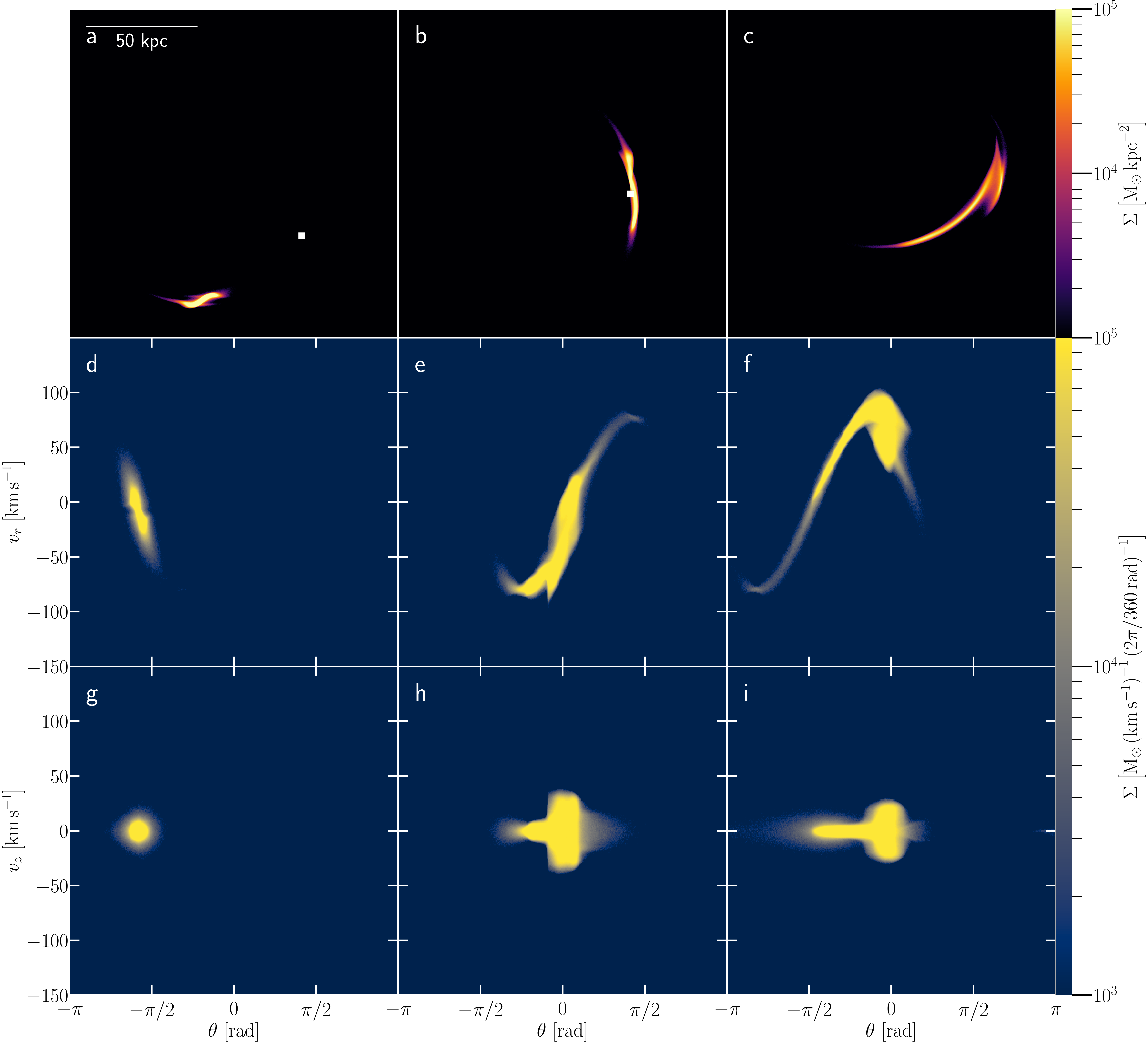}
    \end{center}
    \caption{
    The top row shows the time evolution of the density distribution of the stream and the bottom two rows display the time evolution of the phase-space density distribution with perturbation.
    The white dot represents the mass center of the perturber.
    (a), (d), and (g) correspond to the snapshot at {1.6}~{Gyr}, (b), (e), and (h), {2.1}~{Gyr}, and (c), (f), and (i), {3.0}~{Gyr}.
    The middle row presents the radial velocity as a function of the azimuthal angle, while the bottom row shows the vertical velocity as a function of the azimuthal angle. 
    {Alt text: Nine panels with three times three configuration. In the top row, from left to right, the panels are labeled a to c. In the middle row, from left to right, the panels are labeled d to f.In the bottom row, from left to right, the panels are labeled g to i.}
    }
    \label{fig:phasespace_wp}
\end{figure}

\begin{figure}
    \begin{center}
        \includegraphics[width=8cm]{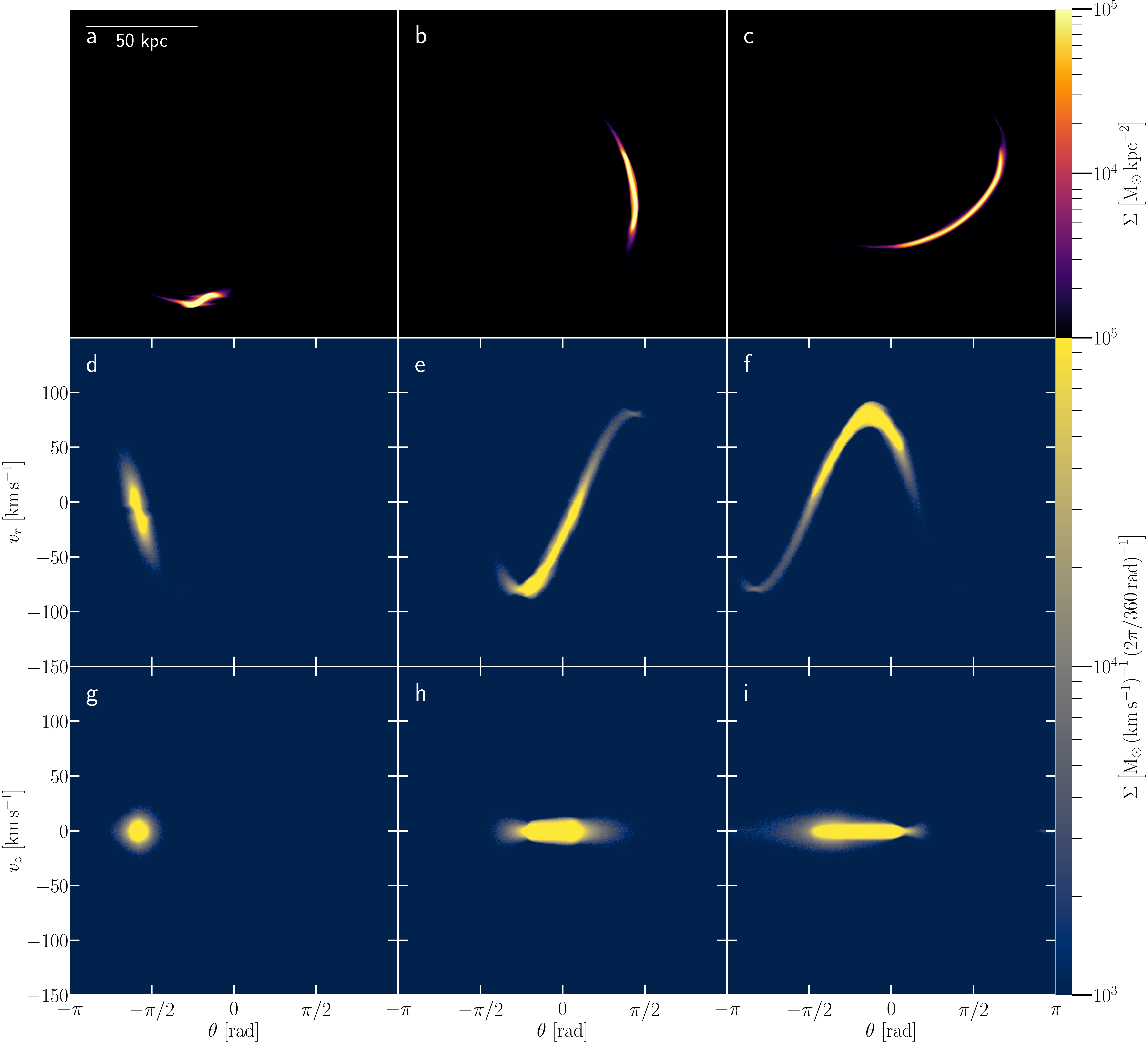}
    \end{center}
    \caption{
    Same as figure 9, but without perturbation.
    {Alt text: Nine panels with three times three configuration. In the top row, from left to right, the panels are labeled a to c. In the middle row, from left to right, the panels are labeled d to f.In the bottom row, from left to right, the panels are labeled g to i.}
    }
    \label{fig:phasespace_wop}
\end{figure}

The time evolution of the projected density of the stream on the orbital plane with the perturber model with mass ${5\times10^8}~{M_\odot}$, size \qty{0.32}{kpc}, and relative velocity ${190}~\mathrm{km\,s^{-1}}$ is shown in the top row of figure~\ref{fig:phasespace_wp}.
The center of gravity of the perturber is shown in the white point.
For comparison, figure~\ref{fig:phasespace_wop} shows the snapshots of the simulation without a perturber.
The interaction between the stream and perturber occurs around \qty{1.5}{Gyr}.
We also show the phase space behavior of the stream in the bottom two rows in figures~\ref{fig:phasespace_wp} and \ref{fig:phasespace_wop}. 
The second row shows radial velocity against azimuthal angle and the third row shows the vertical velocity against azimuthal angle.
Right after the interaction with the perturber, the radial velocity and the vertical velocity of the stream increase toward the positive and negative direction.
Then, the radial velocity change causes the modification of the orbital radius toward the inner and outer directions.
This results in a shift of phase in orbital motion.
Moreover, the tidal effect increases the distance of separated streams.
Therefore, the two--component structure is clear on the orbital plane.
Meanwhile, the scatter in $z$-direction does not enhance the tidal disruption so it calms down soon.
Therefore, the two--component structure becomes most apparent in a face-on view of the orbital plane, where variations in the distance from the host center give rise to the strongest differential tidal forces.

\subsection{Constraint on parameters of the perturber} \label{sec:parameter_survey}
In this section, numerical experiments under a variety of perturber models are presented to explore the factors influencing the splitting of stellar streams. 

\begin{figure}
    \begin{center}
        \includegraphics[width=8cm]{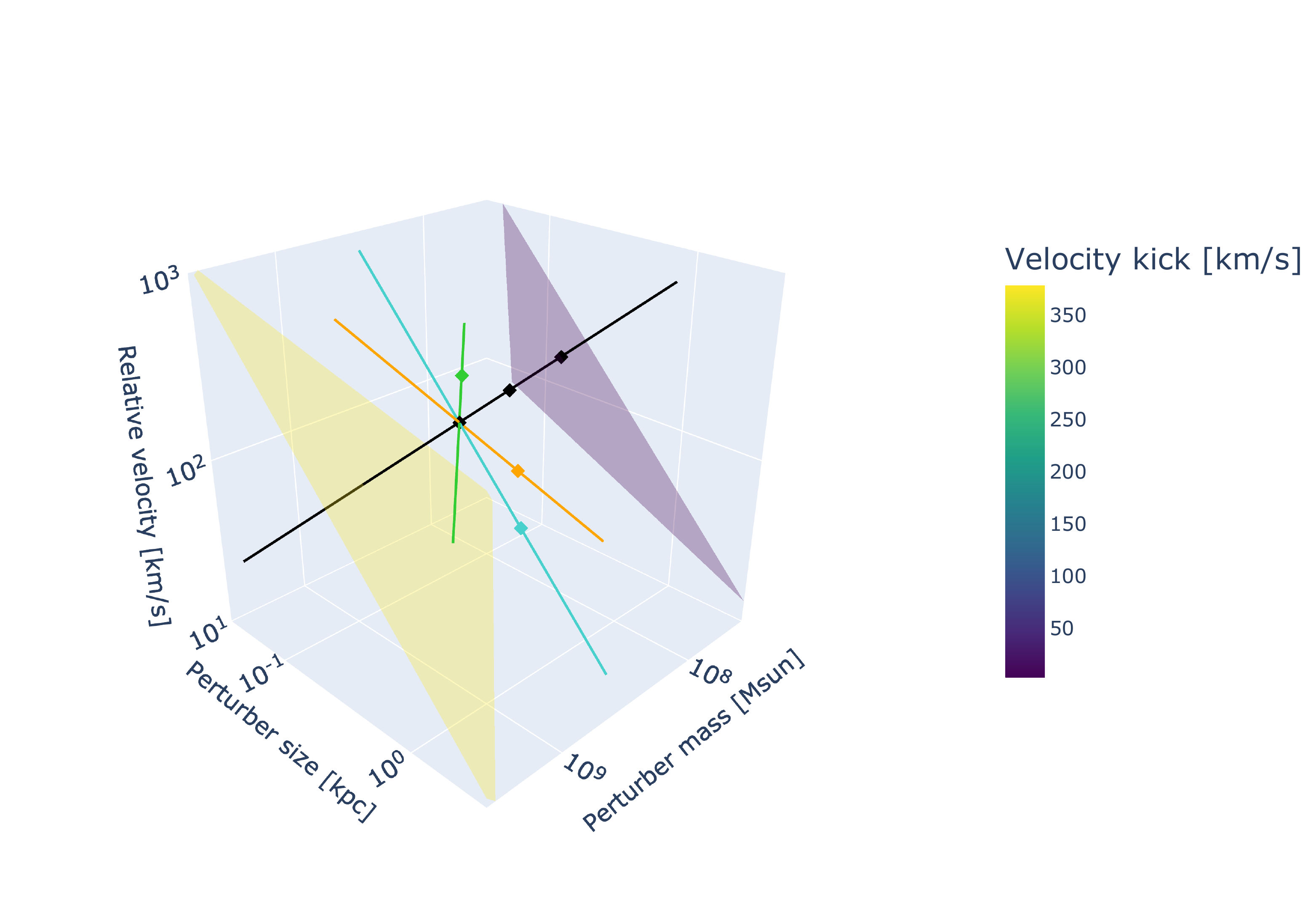}
    \end{center}
    \caption{
    The velocity kick at $b = r_{\rm{per}}$ as a color map on the $M_{\rm{per}}$--$r_{\rm{per}}$--$v_{\rm{rel}}$ space, used as an index to choose the properties of a perturber for parameter survey.
    The violet surface shows the surface where the velocity kick is equal to the initial velocity dispersion of the stellar stream progenitor.
    From model a on the surface, model b and model c are taken in a perpendicular direction (black solid line) to the surface where the velocity kick is equal to the escape velocity of the host dark matter halo (the yellow surface).
    Model d, model e, and model f are taken to have the same velocity kicks as model c.
    Model d, the green point on the green line, has the same $v_{\rm{rel}}$ as model c.
    Model e, the orange point on the orange line, has a same $r_{\rm{per}}$ as model c.
    Model f, the sky blue point on the sky blue line, has a same $M_{\rm{per}}$ as model c.
    {Alt text: Three-dimensional color map of velocity kick amplitude across perturber mass, size, and relative velocity space.}
    }
    \label{fig:parameter_choice}
\end{figure}

To conduct the parameter survey most efficiently, we use the velocity kick imparted by a perturber as an index to choose the properties of a perturber.
The velocity kick is given by the impulse approximation $\Delta v_x = - {2GM_{\rm{per}}b}/\{(b^2+r_{\rm{per}}^2)v_{\rm{rel}}\}$ at $b = r_{\rm{per}}$, the impact parameter with the largest scattering angle.
The velocity kick $\Delta v_x$ at $b = r_{\rm{per}}$ can be expressed as a function of the perturber mass $M_{\rm{per}}$, the perturber size $r_{\rm{per}}$, and the relative velocity $v_{\rm{rel}}$.
Therefore, we can show the velocity kick as a colormap on the $M_{\rm{per}}$--$r_{\rm{per}}$--$v_{\rm{rel}}$ space, as shown in figure~\ref{fig:parameter_choice}.
When we see the variation of the velocity kick on the perturber mass axis, massive perturbers give large velocity kicks to the stream particles.
When we see the variation of the velocity kick on the perturber size axis, smaller perturbers in size exert stronger disturbances due to their steeper gravitational potentials. 
When we see the variation of the velocity kick on the relative velocity axis, slower velocities allow for prolonged interactions, enhancing the gravitational influence of the perturber.

\begin{figure}
    \begin{center}
        \includegraphics[width=8cm]{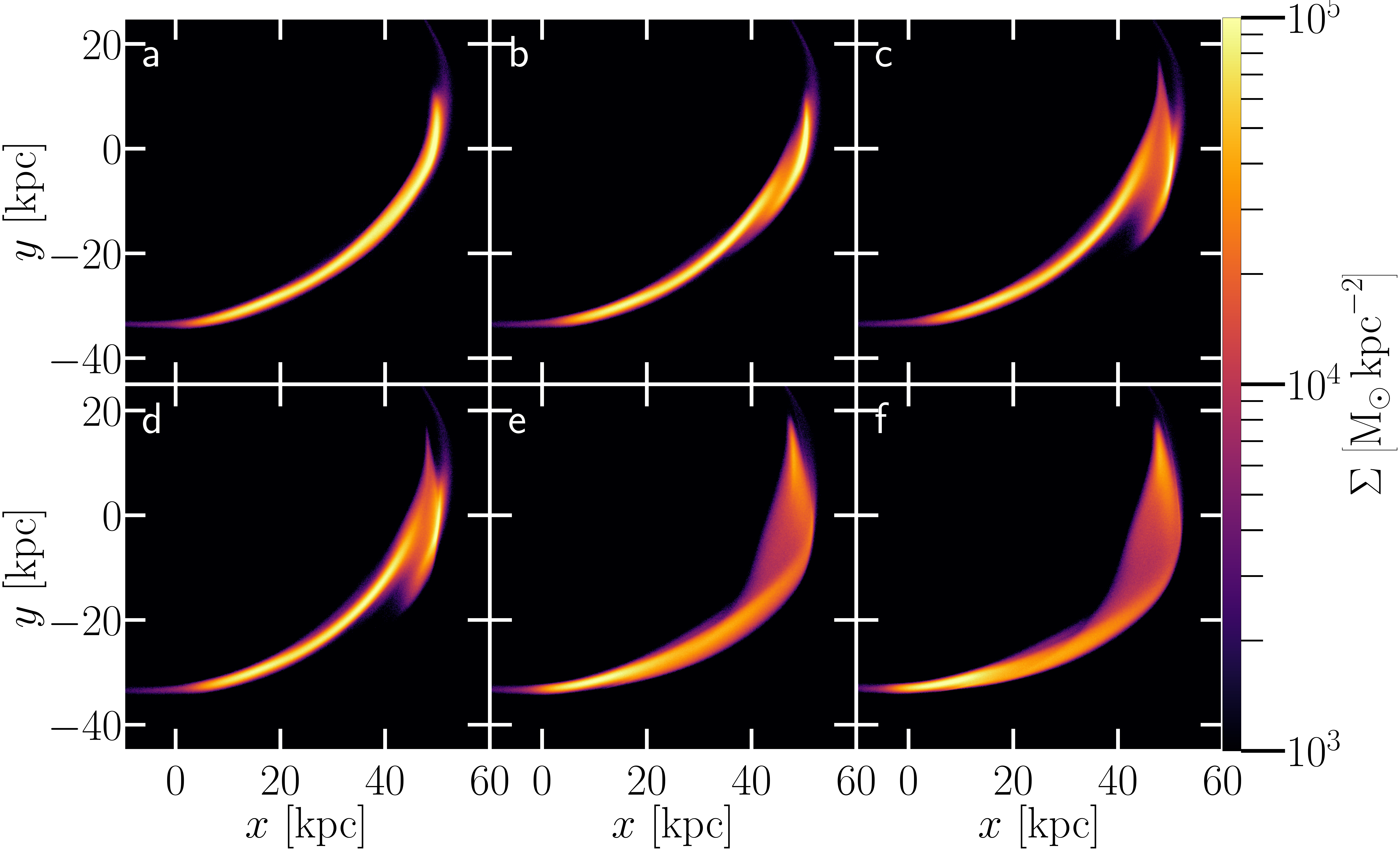}
    \end{center}
    \caption{
    Comparison between different perturber models.
    Simulation snapshots at 3.0~Gyr are shown.
    The properties of each model are listed in Table~\ref{tab:perturber_properties}.
    {Alt text: Six simulation snapshots at 3.0 gigayears for perturber models a through f. Stream splitting increases from model a (weak impulse) to model c (strong impulse). Models d, e, and f show similar morphologies to model c despite different individual parameter values.}
    }
    \label{fig:parameter_survey}
\end{figure}

When the velocity kick equals to the initial velocity dispersion of stellar stream progenitor, one cannot distinguish the perturbation from the stream's own velocity dispersion because perturbation is too small.
Here, we can take the velocity kick constant surface in the $(M_{\rm{per}}$, $r_{\rm{per}}$, $ v_{\rm{rel}})$ space.
The violet surface shows the surface where velocity kick equals to the initial velocity dispersion of stellar stream progenitor $\sim 20~\mathrm{km\,s^{-1}}$.
Perturbers that cannot impart a velocity kick larger than the initial velocity dispersion of the stream, upper right side of the violet surface, are insufficient to split a stellar stream.
We take model a on the surface where velocity kick equals to the initial velocity dispersion of stellar stream progenitor.
The snapshots of the simulations after the interactions for model a is shown in figure~\ref{fig:parameter_survey}(a).
As seen from the figure, one cannot confirm the changes when the perturbation is comparable to the progenitor's velocity dispersion.

Conversely, perturbers that impart a velocity kick larger than the escape velocity of the host dark matter halo $\sim 370~\mathrm{km\,s^{-1}}$, the lower left side of the yellow surface, the stream is destroyed by the interaction.
Along the black solid line, which is perpendicular to the velocity-kick-constant surface, the velocity kick increases from model b to model c.
Model c adopts the same parameters as the simulation presented in the previous section.  
However, we note that the perturber is modeled as a Plummer sphere in this work to adopt a general and simplified density profile, and therefore does not strictly follow the concentration--mass relation.
The snapshots of the simulations for model b and model c are shown in figures~\ref{fig:parameter_survey}(b) and (c), respectively.
As seen from the figures~\ref{fig:parameter_survey}(a), (b), and (c), the larger the impulse, the stronger the perturbation.
When calculating the one-dimensional density profile along the direction perpendicular to the stream elongation, we define the spread width as the distance between points where the density gradients become significant, and the separation width as the distance between density peaks. For model a, the spread width is $8.6~\mathrm{kpc}$, while no separation is detected. For model b, the spread width and separation width are $11.6~\mathrm{kpc}$ and $2.9~\mathrm{kpc}$, respectively. For model c, the spread width and separation width are $12.2~\mathrm{kpc}$ and $5.2~\mathrm{kpc}$, respectively.

When the perturber properties come closer to the surface with the velocity kick equals to the escape velocity of the host dark matter halo, we confirm the destruction of the coherent motions of the stream particles in the simulation.
However, as the stream particles scattered by large angle in this regime, we further need to confirm the results with higher order time integration scheme with direct $N$-body calculation.

Moreover, we took three more models that have the same velocity kicks as the model c, but different properties, namely, $M_{\rm{per}}$, $r_{\rm{per}}$, and $v_{\rm{rel}}$.
To make the comparison clear, we keep one parameter of perturber constant and change the other two parameters in $(M_{\rm{per}}, r_{\rm{per}}, v_{\rm{rel}})$ space.
Model d, the green point on the green line, has the same $v_{\rm{rel}}$ as model c but different $r_{\rm{per}}$ and $M_{\rm{per}}$.
The MSE of the density distributions of model c and model d is small enough to conclude that they show pretty similar structures. 
This means that the $M_{\rm{per}}$ and $r_{\rm{per}}$ are degenerated.
Model e, the orange point on the orange line, has the same $r_{\rm{per}}$ as model c but a different velocity and $M_{\rm{per}}$.
model f, the sky blue point on the sky blue line, has the same $M_{\rm{per}}$ as model c but different $r_{\rm{per}}$ and $v_{\rm{rel}}$.
Since model e and model f are at the same distance from model c in $(M_{\rm{per}}, r_{\rm{per}}, v_{\rm{rel}})$ space, the $v_{\rm{rel}}$ of model e and model f is the same.
From the comparison between figure~\ref{fig:parameter_survey}(e) and (f), we can also conclude that the $M_{\rm{per}}$ and $r_{\rm{per}}$ are degenerate.
And the velocity makes change in the morphology of split.
The properties of perturber models used in our analysis are summarized in Table~\ref{tab:perturber_properties}.
The parameter survey using simulations show that stream splitting occurs only when the parameters lie between the surface where velocity kick equals to the initial velocity dispersion of the stellar stream progenitor (the violet surface) and the surface with the velocity equals to the escape velocity of the host dark matter halo (the yellow surface).
As an illustrative estimate, adopting a relative velocity of $200~\mathrm{km\,s^{-1}}$ together with the MW subhalo concentration--mass relation from \citet{Kaneda2024}, the parameter 
region capable of producing stream splitting corresponds to
$9.1\times10^{5}~M_{\odot} < M_{\mathrm{per}} < 1.3\times10^{9}~M_{\odot}$.

The trends reported here should be regarded as order-of-magnitude trends within the explored parameter range. A denser and wider parameter survey may reveal additional factor-level dependencies, which are beyond the scope of the present work.

\begin{table}
    \centering
    \caption{
    The properties of perturber models.
    The first column shows the label of the perturber models. The second column shows mass of perturber.
    The third column shows size, namely, core radius of Plummer sphere, of the perturber.
    The fourth column shows relative velocity between the stellar stream and the perturber. 
    }
    \label{tab:perturber_properties}
    \begin{tabular}{lccc}
    \hline
    Model & mass & size & relative velocity\\
    \hline
    a & ${2.2\times10^8}~{M_\odot}$ & ${0.71}~\mathrm{kpc}$ & ${430}~\mathrm{km\,s^{-1}}$\\
    b & ${3.3\times10^8}~{M_\odot}$ & ${0.48}~\mathrm{kpc}$ & ${290}~\mathrm{km\,s^{-1}}$\\
    c & ${5\times10^8}~{M_\odot}$ & ${0.32}~\mathrm{kpc}$ & ${190}~\mathrm{km\,s^{-1}}$\\
    d & ${3\times10^8}~{M_\odot}$ & ${0.20}~\mathrm{kpc}$ & ${190}~\mathrm{km\,s^{-1}}$\\
    e & ${3\times10^8}~{M_\odot}$ & ${0.32}~\mathrm{kpc}$ & ${120}~\mathrm{km\,s^{-1}}$\\
    f & ${5\times10^8}~{M_\odot}$ & ${0.52}~\mathrm{kpc}$ & ${310}~\mathrm{km\,s^{-1}}$\\
    \hline
    \end{tabular}
\end{table}

\begin{figure}
	\begin{center}
	\includegraphics[width=8cm]{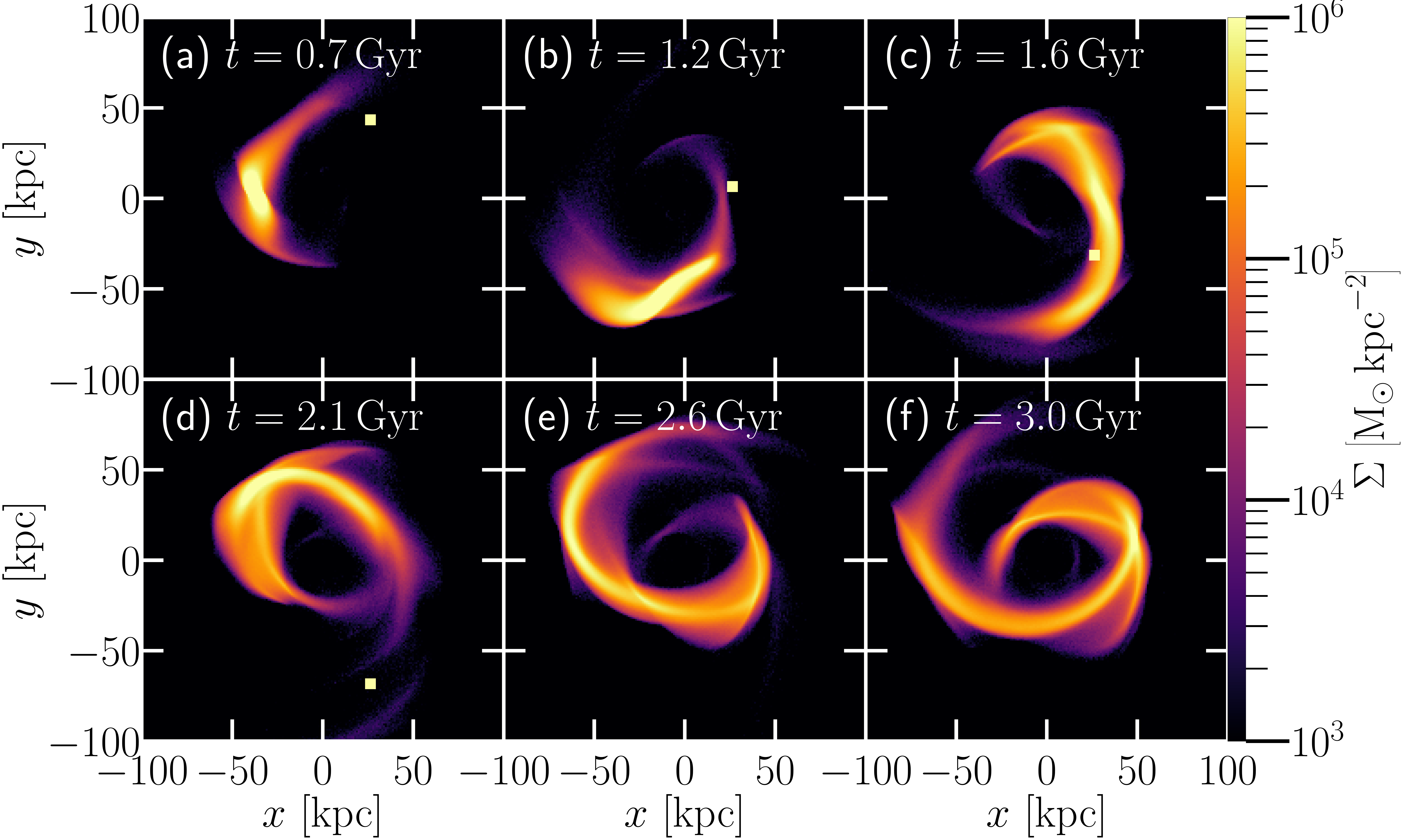}
    \end{center}
	\caption{   
        The time evolution of the face-on stream density to the orbital plane.
        The panels labeled a to f, show the time evolution of the simulation after the collision between a starless dark matter subhalo and a stellar stream.
        The surface mass density of the stream is shown in the color map, with the mass center of the starless dark matter subhalo indicated by a yellow dot. In this model, the host dark matter halo has a mass of $\SI{e+12}{M_\odot}$. 
        The stream has a total mass of ${10^9}{M_\odot}$, and the colliding dark satellite has a mass of ${5\times 10^9}{M_\odot}$, moving with a relative velocity of $\qty{250}{km.s^{-1}}$.
        The time elapsed since the start of the calculation is represented in each panel as follows: (\textbf{a}) 0.7 Gyr, (\textbf{b}) 1.2 Gyr, (\textbf{c}) 1.6 Gyr, (\textbf{d}) 2.1 Gyr, (\textbf{e}) 2.6 Gyr, and (\textbf{f}) 3.0 Gyr.
        {Alt text: Six-panel time sequence from 0.7 to 3.0 gigayears showing face-on density of a massive stellar stream encountering a dark matter subhalo, confirming parallel stream formation at larger mass scales.}
        }
	  \label{fig:timeevolution_massive}
\end{figure}

Furthermore, we simulate an interaction involving a more massive stellar stream and a starless dark matter subhalo. 
The time evolutions of density maps are presented in figure~\ref{fig:timeevolution_massive}. In this setup, the stellar stream possesses a total mass of ${10^9}~{M_\odot}$, while the impacting dark subhalo has a mass of ${5\times10^9}~{M_\odot}$ and approaches with a relative velocity of ${250}~\mathrm{km~s^{-1}}$. The six panels (a–f) depict the temporal evolution of the stellar stream following its encounter with the starless dark matter subhalo. The colour map indicates the surface mass density, and the yellow dot marks the subhalo’s center of mass. The elapsed times since the beginning of the simulation are: (a)~0.7~Gyr, (b)~1.2~Gyr, (c)~1.6~Gyr, (d)~2.1~Gyr, (e)~2.6~Gyr, and (f)~3.0~Gyr.
We confirm that parallel stellar streams can form across different mass scales within a MW--sized host halo.

\section{Discussions}
\subsection{Discrimination from non-perturbation formation scenarios}
\begin{figure*}
    \begin{center}
        \includegraphics[width=\textwidth]{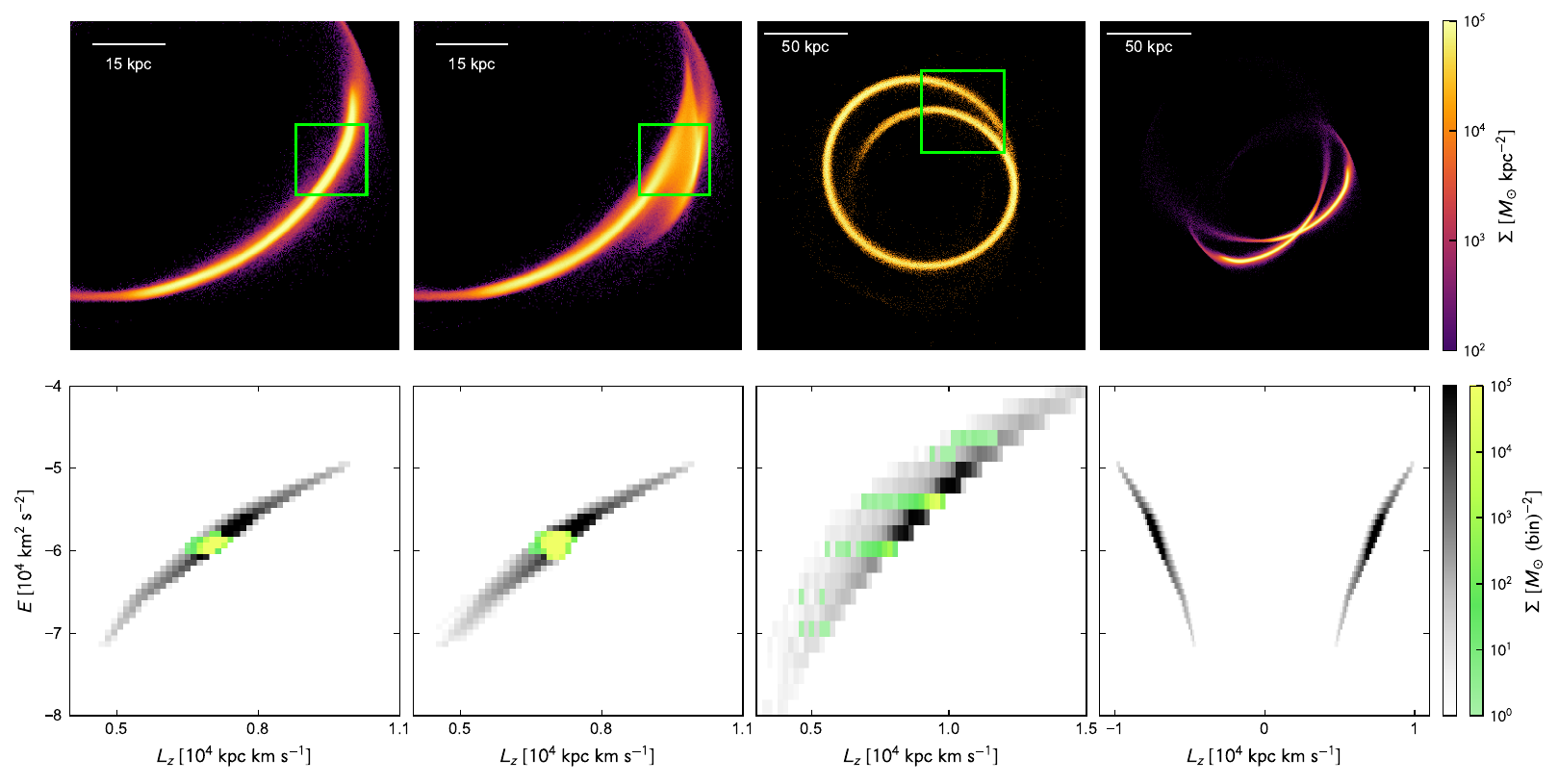}
    \end{center}
    \caption{
    The lower row panels display the density distribution of the stream on the $E-L_z$ plane, based on the simulation results.
    The top row panels show the corresponding spatial density distribution of the stream of the face on view to the orbital plane. 
    The left most column displays the unperturbed case, while the second left column shows the perturbed case. 
    The third left column shows the extended stream that has overlap with its own tail, also looks like parallel structure.
    The right most column shows the two distinct streams lies along in the space.
    In the bottom panels, the density distributions of the entire streams are shown in grayscale.
    For x-axis, one bin corresponds to $\sim {100}~\mathrm{kpc}~\mathrm{km\,s^{-1}}$ and for y-axis, one bin corresponds to $\sim {1000}~\mathrm{km^2\,s^{-2}}$.
    Overlaid in the color map are the density distributions of the particles located within the regions that show parallel structures.
    Corresponding spatial regions are highlighted by the green squares in the upper panels.
    {Alt text: Four columns, two rows. Top row: face-on spatial density maps for unperturbed, perturbed, self-overlapping, and two-distinct-stream scenarios, with parallel regions highlighted in green. Bottom row: corresponding density distributions on the energy versus angular momentum plane, showing distinct positions enabling scenario discrimination.}
    }
    \label{fig:E-Lz}
\end{figure*}

In the Gaia mission, an exceptionally large number of stellar streams were discovered \citep{Malhan2018, Malhan2019_phasespace, Malhan2022_atlas}. 
Thanks to the Gaia mission, we now have access not only to the positions of member stars in 
stellar streams on the sky plane, but also their distances from the solar system, line-of-sight velocities, and proper motions are available.
Therefore, we can derive the energy and angular momentum of the stream from observations.
Figure~\ref{fig:E-Lz} shows simulated stellar streams, with the upper panels displaying their spatial distribution and the lower panels illustrating their structure in the energy–-angular momentum ($z$-direction) plane, rendered as grayscale and color maps.
The left most column show the same snapshot without a perturber, for the reference.
The second left column illustrate the simulation snapshot at 3.0 Gyr after the start of the simulation with a perturber model of $5\times10^8~M_\odot$ in mass, 0.32 kpc in size, $190~\mathrm{km~s^{-1}}$ in relative velocity.
The density distribution in the energy--angular momentum plane of the particles in the parallel part enclosed by the green square in the upper panels are shown by the colored bars in the lower panels.
The stellar particles in the parallel region 
appears in a single side of the extended stream distribution on the energy--angular momentum plane.
One stream stretched more than a lap are shown in the third left column.
The stellar particles in the parallel region appears on both sides of the stretched energy--angular momentum distribution and does not appear as a single side.
Therefore, one can distinguish a streached and overlapped stream with a stream split by a perturbation using the stream distribution on the energy--angular momentum plane.
Moreover, the right most panels show the streams of different origins that are spatially adjacent.
On the energy--angular momentum plane, these streams occupy different positions.
Therefore, we can discern different origin streams from one origin stream split with perturbation on the energy--angular momentum plane.
In the bottom panels of figure \ref{fig:E-Lz}, one bin corresponds to $\sim {100}~\mathrm{kpc}~\mathrm{km\,s^{-1}}$ for the angular momentum in $z$ direction and one bin corresponds to $\sim {1000}~\mathrm{km^2\,s^{-2}}$ for the energy.
This is in the same order as the observational error in the analysis provided in \citet{Malhan2022_atlas} based on Gaia Early Data Release 3 (see Table 2 of \cite{Malhan2022_atlas}).
For example, Indus and Jhelum, which are parallel to each other on the sky plane, are proposed to have the same origin since they share similar width, heliocentric distance, and metallicity \citep{Bonaca2019_multiple, shipp2018}.
Moreover, they share a close position on the energy-momentum plane (see figures~1 and 2, and Table 2 in \cite{Malhan2022_atlas}).
A diagnosis based on observational properties of streams such as orbital parameters combined with sky position and metallicities will provide further supporting evidence for the formation process of parallel streams and the existence of dark perturbers. 

\subsection{Other Scenarios for Perturbing Streams}
One interpretation for density clumps in the globular cluster streams are thought to be the result of longitudinal Jeans instabilities \citep{quillen_jeans_2010, comparetta_simulations_2010}.
After that, \citet{schneider_stability_2011} updated their model, taking the diffluence of the stream stars into account.
They confirmed that globular cluster streams are stable for the entire stream and over all time scales.

The stellar density variations in the stream can also be due to known baryonic structures such as giant molecular clouds, globular clusters, dwarf galaxies, or the MW’s bar or spiral arms when they are close to the host galaxy center \citep{amorisco_gaps_2016, erkal_sharper_2017, pearson_gaps_2017}.
In the Stream C and Stream D case, they are unlikely to be perturbed by the baryonic component of M31 as the Stream C and Stream D are $\sim 60 ~\mathrm{kpc}$ from the M31 center, far away from the baryonic component of M31 ($\lesssim 15 ~\mathrm{kpc}$).
The Jhelum and Indus streams are $\sim 20 ~\mathrm{kpc}$ from the MW center.
In a vertical distance, the Jhelum and Indus streams lie at $z \sim 10$--$15~\mathrm{kpc}$ from the Galactic disk, making it extremely unlikely to be affected by disk components (bar, spiral arm, giant molecular clouds) confined within $\lesssim 100~\mathrm{pc}$.
However, globular clusters and dwarf galaxies are distributed over $15 ~\mathrm{kpc}$.
A possible next step is to integrate the perturber orbit backward in time and assess whether any known globular cluster or dwarf galaxy with a measured trajectory has an orbit consistent with the required encounter.

A detailed investigation of host-halo asymmetry is beyond the scope of this work. 
Nevertheless, we note that a non-spherically symmetric host dark matter halo potential can also affect the morphology of stellar streams \citep{yavetz_stream_2023}.

\subsection{Variety in morphology of perturbed stream}
There is a variety in the morphology of perturbed stellar streams, depending both on the properties of the stream progenitors and the perturbers.

In this study, we assume that a stream is fully disrupted by the tidal field of the host galaxy. 
However, pre-impact stream conditions can yield a wider set of morphological imprints. 
If the stream retained a progenitor nucleus before the collision and is fully disrupted afterward, stars in that nucleus possibly produce a thin, high-density component alongside split tails by a collision. This may explain the coexistence of a narrow component (scattered nucleus) and a broad component (split tails) in the Jhelum stream. By contrast, if the progenitor nucleus survives the collision, it may produce bifurcated streams linked to a central core, as in the Sagittarius stream. 

Gaps, density depletion along the stream, are thought to be produced by subhalo flyby, and actively investigated in previous studies.
The primary differences between cases where gaps form in streams and those where parallel streams form are the angle of the relative velocity as seen from the stream, which determines the direction of the imparted velocity kick, and the duration over which a perturber influences the stream. A perpendicular collision imparts a velocity kick in the parallel direction and cuts the stream, whereas a parallel collision imparts a velocity kick in the perpendicular direction and splits the stream into two parallel streams.
The instantaneous and localized impact produces the gaps in stellar streams, while the starless dark matter subhalo crossing in the stellar stream over a certain period tears a stream into two parallel streams. 
Our framework, which focuses on the density variation in the perpendicular direction along the stream, enables a more comprehensive characterization of stream–subhalo interactions.
Extending the framework to more general oblique encounters will be investigated in a forthcoming study.

The pre-encounter structure of the stream and the orbit of the perturber may give rise to a wider variety of signatures, such as kinks \citep{Li2021}, spurs \citep{Bonaca2019}. 
A systematic investigation of these morphologies will be important for deriving statistically robust constraints on the abundance of dark perturbers.

\subsection{Expected Encounter Rate in MW-scale Galaxies}
As a first step toward assessing whether such signatures can occur in MW-scale galaxies, we estimate the encounter rate between stellar streams and dark perturbers.
Collision frequency between globular cluster streams and dark matter subhalos are investigated in the literatures \citep{yoon_clumpy_2011, Carlberg2012, banik_novel_2021, banik_evidence_2021}.
Therefore, here we focus on estimating (1) the collision frequency between dwarf galaxy streams and subhalos, and (2) the collision frequency between globular cluster streams and IMBHs.

\subsubsection{Collision Frequency between Dwarf Galaxy Streams and Subhalos}
We estimate the frequency of collisions between subhalos with galaxies
and subhalos without galaxies, to assess the number of potential observational signatures.
The collision frequencies between subhalos within the sum of their scale radius, as derived from the average of cosmological simulations, is $2.1 \times 10^2 ~\mathrm{Gyr}^{-1}$ \citep{OtakiKazunoMori2025}.
They also provide an estimate of the collision fraction between subhalos with galaxies and subhalos without galaxies to be 0.35, based on observations of dwarf galaxy--dwarf galaxy interactions.
Therefore, the collision frequency of subhalos with galaxies and subhalos without galaxies is $73.5 ~\mathrm{Gyr}^{-1}$.
If we assume collisions with the encounter angle between the stellar stream and the perturber smaller than $30\degree$ result in the formation of parallel--like structures, approximately 10\% of all collisions produce parallel-like stellar streams.
When the streams are destroyed during the dynamical relaxation of their constituent particles, the observable timescale could extend to several hundred Myr. 
Consequently, the expected encounter rate is not obviously inconsistent with the presence of several observable remnants in the M31.

\subsubsection{Collision Frequency between Globular Cluster Streams and IMBHs}
We then focus on estimating the frequency of collisions between globular cluster streams and IMBHs.
In Sec. \ref{sec:parameter_survey}, we show that perturbations are indistinguishable from the intrinsic velocity dispersion of the stellar stream when the imparted velocity kick is smaller than the initial velocity dispersion of the progenitor system. 
Thereby, while the relatively small masses of IMBHs considered here are insufficient to produce detectable perturbations for dwarf galaxy streams, they are sufficient for globular cluster streams.
Therefore, we only focus on interactions between IMBHs and globular cluster streams here.

Several origins of IMBHs have been proposed \citep{Greene2020}, affecting their abundance and mass range. 
Remnants of massive first-generation stars are considered the dominant IMBH population in the MW halo \citep{RashkovMadau2014}, though their growth is limited unless absorbed into larger subhalos. 
IMBHs formed through runaway collisions in star clusters are also expected to wander due to gravitational recoils \citep{Holley-Bockelmann2008}. 
In addition, tidal disruption of infalling satellites may produce a population of wandering IMBHs within galactic halos.
Considering this scenario, a few tens of such black holes ($>{10^4}~{M_{\odot}}$) should be wandering in the MW-sized halo \citep{RashkovMadau2014}. 
A part of IMBHs has been recoiled through three-body encounters of massive black holes, surrounded by very compact star clusters ($<{1}~\mathrm{pc}$) \citep{O'LearyLoeb2009,O'LearyLoeb2012}. 
They could also behave as very compact objects that mimic IMBHs of $\sim {10^5}~{M_{\odot}}$. 
Although the number of each target is not large, they are distributed within $\sim {30}~\mathrm{kpc}$ from the host center due to the necessity of tidal disruption of the parent dwarf galaxy. 
Cosmological studies estimate $\sim 10$ IMBHs in MW--sized halos, though large-scale simulations often seed black holes of $10^{5-6}~\mathrm{M_\odot}$ due to resolution limits \citep{Weller2022}. 
Seeding in $> {10^8}~{M_\odot}$ halos predicts $\sim 5$ IMBHs within {10}~{kpc} and $\sim 10$ in the inner halo today \citep{Tremmel2018}. 
Similar estimates are supported by semi-analytical models \citep{Untzaga2024}. 
Alternative seeding methods, such as in collapsing gas clumps, predict 15 wandering IMBHs in MW--like halos \citep{Ricarte2021}. 
Comparisons of different simulation techniques suggest {5}--{18} IMBHs may wander in MW--sized galaxies, even at $z \geq 3$ \citep{vanDonkelaar2025}. Based on these findings, we adopt an estimate of 20 IMBHs for this analysis.

A wide-field survey of the M31 has revealed that more than half of the globular clusters in the outer halo region are associated with tidal substructures formed by disrupted dwarf galaxies \citep{Veljanoski2014,Veljanoski2015,Mackey2019}.
Theoretical studies focusing on globular cluster streams are still in progress \citep{Pearson2024}. 
A study suggests that halos with masses $\gtrsim {10^9}~{M_\odot}$ in the early stages of galaxy formation, each hosting three globular clusters, could produce {20}--{30} streams within {50}~{kpc} of a MW--sized halo at present \citep{Carlberg2018}. 
Gaia observations have identified long stellar streams in 15 globular clusters \citep{Ibata2021}, while dozens of others show tidal features \citep{Piatti2020,Kuzma2025}. Pericentric distances of globular clusters are estimated to be $\lesssim {10}~\mathrm{kpc}$ \citep{Malhan2022_atlas}. 
Notable examples include Palomar~5 and NGC~5466, observed as the cores of stellar streams \citep{GrillmairDionatos2006_Pal5,GrillmairJohnson2006,Koposov2010,Bowden2015,Bovy2016}. The GD-1 stream, likely formed by a globular cluster, exhibits a dynamically cold nature, with a pericentric distance of {14.1}~{kpc} \citep{GrillmairDionatos2006_GD1,Bonaca2019,MalhanIbata2019}. These findings suggest many faint, undiscovered streams may exist around inner halo globular clusters. 
Recent numerical studies indicate that while numerous globular clusters formed early in MW--like halos, most (80--90 per cent) lost their mass over time \citep{Baumgardt2022,Grudic2022,Rodriguez2023,Chen2024}. Some contributed to the stellar halo of the host galaxy, but many may still form cluster streams, exceeding current expectations.
Therefore, in addition to the $\sim$30 globular cluster streams, we account for 140 globular clusters distributed in the inner halo \citep{Harris2010,Baumgardt2019,Vasiliev2019} as potential sources of faint tidal streams in the MW. 

Assuming a random process, the collision frequency between globular cluster streams and IMBHs can be estimated as:
\begin{equation}
\frac{\mathrm{d}{F}}{\mathrm{d}t} = \frac{\sigma_{\mathrm{str}} n_{\mathrm{str}}  n_{\mathrm{bh}}  v_{\mathrm{rel}}}{V_{\mathrm{sys}}},
\end{equation}
where $n_{\mathrm{str}}$ and $n_{\mathrm{bh}}$ are the numbers of globular clusters and IMBHs, respectively, $v_{\mathrm{rel}}$ is their relative velocity, and $V_{\mathrm{sys}}$ is the considered volume. 
The collision cross-section is given by $\sigma_{\mathrm{str}} \sim  L_{\mathrm{str}}w_{\mathrm{str}}$, where $L_{\mathrm{stream}} = 10~\mathrm{kpc}$, and $w_{\mathrm{stream}} = 0.1~\mathrm{kpc}$. 
A typical relative velocity of $v_{\mathrm{rel}} = {440}~{\mathrm{km\,s}^{-1}}$ is assumed, corresponding to twice the MW’s circular velocity. 
We consider the volume of inner halo within $25~\mathrm{kpc}$, $V_{\mathrm{sys}} = \frac{4}{3}\pi(25~\mathrm{kpc})^3$.
Setting $n_\mathrm{gc}=170$ and $n_\mathrm{bh}=20$, the estimated collision frequency is approximately ${\mathrm{d}{F}}/{\mathrm{d}t} \sim 2.3~\mathrm{Gyr}^{-1}$ in the MW. 
In the M31, over 300 globular clusters have been identified \citep{BarmbyHuchra2001,Perrett2002,Galleti2009,CaldwellRomanowsky2016}, leading to a higher collision frequency of ${\mathrm{d}{F}}/{\mathrm{d}t} \sim 4.1~\mathrm{Gyr}^{-1}$. The number density of IMBHs increases closer to the halo center \citep{Tremmel2018}, suggesting that the collision frequency may rise in the central regions of the MW.

If we assume collisions with the encounter angle between the stellar stream and the perturber smaller than $30\degree$ result in the formation of parallel--like structures, approximately 10\% of all collisions produce parallel-like stellar streams.
Assuming the observable timescale could extend to several hundred Myr, it would not be unexpected for remnants of such encounters to be present in the MW.

\section{Summary}
In this work, we propose a new dynamical signature, parallel stellar streams, as imprints left by perturbations from starless dark matter subhalos or IMBHs in the MW and M31.
We summarize our main findings as:
\begin{enumerate}
    \item The analytic model and simulations confirm that a single stream can split into two parallel streams.
    \item $N$-body simulations show that the perturber mass, size, and relative velocity must be in a certain region to tear the stream.    
    In other words, these results indicate that the properties of the perturber might be inferred from detailed stream morphologies.    
    For example, assuming the relative velocity between a dark matter subhalo and a stellar stream comparable to the circular velocity of the host halo $v_\mathrm{rel} \simeq 200~\mathrm{km\, s^{-1}}$ and adopting the median $c$--$M$ relation for the subhalo density profile, the mass of a perturber need to be from $9.1 \times 10^{5} \, M_\odot$ to $1.3 \times 10^{9} \, M_\odot$. 
    To produce a split with a characteristic separation of $\simeq 15~\mathrm{kpc}$, as observed in Stream C and Stream D, it requires a mass of perturber $\simeq 5\times10^8~M_\odot$.
    \item The energy--angular momentum plane, which can be derived from Gaia satellite data, can be used as a diagnostic for discerning the formation scenario of the parallel stellar streams.
    \item Based on order-of-magnitude estimates, the expected encounter rate is not obviously inconsistent with the presence of a small number of observable remnants in MW-- or M31--like galaxies.
\end{enumerate}
To conclude, our results demonstrate that parallel stellar streams can arise naturally from stream--perturber encounters and should be considered as part of the broader morphological response of stellar streams to dark perturbers.

\begin{ack}
We are grateful to Yudai Kazuno for providing simulation results of subhalo collisions, as well as to Tomoaki Ishiyama for providing simulation results of cosmological $N$-body simulation.
We also thank Toshihiro Kawaguchi for his valuable comments and constructive feedback.
This research (in part) used computational resources of Wisteria/BDEC-01 Odyssey (Information Technology Center, The University of Tokyo) and Miyabi (Joint Center for Advanced High Performance Computing (JCAHPC)), provided by the Multidisciplinary Cooperative Research Program in the Center for Computational Sciences, University of Tsukuba.
List the grants, fellowships etc. that funded the research;
YK was funded by JSPS KAKENHI Grant Numbers JP23KJ0280, 26KJ0085, and JSPS International Leading Research grant 22K21349.
MM was funded by JP24K07085 and JP24K00669.
YM was funded by JP20K14517 and JP23K11123.
TK was funded by JP22K14076 and JP26K07133.
\end{ack}

\bibliographystyle{pasj}
\bibliography{ref}

\end{document}